%% file: manuscript_cs_simulations7.tex
\documentclass[twocolumn, times]{aastex63}

\usepackage{amsmath,verbatim}
\usepackage{xcolor}
\usepackage{natbib}
\usepackage{enumitem}
\usepackage{enumerate}
\setlist[itemize]{leftmargin=*}
\setlist[enumerate]{leftmargin=*}
\received{August 24, 2020}
\revised{March 28, 2021}
\accepted{March 31, 2021}
\newcommand{\us}{\ensuremath\mathrm{\mu s}} %microsecond
\shorttitle{Deconvolution with CS}
\shortauthors{Dolch et al.}
\newcommand{\logten}{\mathop{\log_{10}}}

\begin{document}

\setlength{\abovedisplayskip}{4pt}
\setlength{\belowdisplayskip}{4pt}

%\apj
\title{{\refo Deconvolving Pulsar Signals with Cyclic Spectroscopy: A Systematic Evaluation}}

\input{authors.tex}

\input{abstract.tex}

\keywords{ 
Gravitational Waves ---
Pulsars ---
ISM --- 
}

%\abovedisplayskip
%\belowdisplayskip

%%%%%%%%%%%%%%%%%%%%%%%%%%%%%%%%%%%%%%%%%%%%%%%%%%%%%%%%%%%%%%%%%%%%%%%%%%%
% How to do references using BibTex and ADS:
%
% It is reasonably convenient to use bibtex and natbib.sty
% to format the citations, and to use the bibcodes from ADS. 
% The python script get_bibtex.py, run from the makefile, will scrape the ADS
% for the full citation and put it in the .bib file. Example citations:
% 
% \citep{1989ApJ...337...34Z, 1994, ApJ...435..540C} % surrounds with parentheses
% \citet{1989ApJ...337...34Z, 1994} % surrounds only the year in parenthesies
% \citealt{1989ApJ...337...34Z, 1994} % no parentheses
%  
% See natbib.sty for more info
%%%%%%%%%%%%%%%%%%%%%%%%%%%%%%%%%%%%%%%%%%%%%%%%%%%%%%%%%%%%%%%%%%%%%%%%%%%

%%%%%%%%%%%%%%%%%%%%%%%%%%%%%%%%%%%%%%%%%%%%%%%%%%%%%%%%%%%%%%%%%%%%%%%%%%%
%% How to do figures interspersed with the text for emulateapj mode 
% 
%\ifsubmode
%\else
%\begin{inlinefigure}
%\begin{center}
%\resizebox{\textwidth}{!}{\includegraphics{figures/fig1.pdf}}
%\figcaption{ \figonecap }
%\end{center}
%\end{inlinefigure}
%\fi

%%%%%%%%%%%%%%%%%%%%%%%%%%%%%%%%%%%%%%%%%%%%%%%%%%%%%%%%%%%%%%%%%%%%%%%%%%%
%\setcounter{section}{0}
\section{Introduction \label{sec:intro}}
Millisecond pulsars (MSPs) are excellent tools for long-period gravitational wave (GW) detection. GWs are perturbations in the space-time metric predicted by general relativity, first indirectly detected in a binary neutron star \citep{1975ApJ...195L..51H} and then directly by the Laser Interferometer Gravitational-Wave Observatory (LIGO; \citealt{2016PhRvL.116f1102A}). Using pulsar timing for GW detection requires measurements of time-of-arrival (TOA) perturbations from many MSPs. Long-term monitoring of MSPs distributed widely across the sky (pulsar timing arrays; PTAs) is a robust method for {\refw nanohertz (i.e., light-year wavelength) GW detection} (\citealt{1978SvA....22...36S}, \citealt{1990ApJ...361..300F}). The PTA method of {\refb detecting GWs with frequencies in the range} {\refbfr 10$^{-9}$ to 10$^{-7}$\,Hz} is complementary to {\refxxx laser interferometers}, which probe GW  frequencies of $>$\,10$^{-6}$\,Hz. {\refw PTAs are sensitive to GWs because metric disturbances induce a variation of the observed {\refxxx pulse arrival time}; hence, the TOA will be delayed or advanced relative to a predicted arrival time. PTA collaborations include} the North American Nanohertz Observatory for Gravitational Waves (NANOGrav; \citealt{2013CQGra..30v4008M}), the EPTA (the European Pulsar Timing Array; \citealt{2013CQGra..30v4009K}), and the PPTA (the Parkes Pulsar Timing Array; \citealt{2013CQGra..30v4007H}, \citealt{2013PASA...30...17M}), working collectively as the IPTA (International Pulsar Timing Array; \citealt{2010CQGra..27h4013H}, \citealt{2013CQGra..30v4010M}). {\refz The GW sources to which PTAs are likely the most sensitive are} merging supermassive black hole binaries (\citealt{1979ApJ...234.1100D}, \citealt{1983ApJ...265L..39H}, {\refz \citealt{2018ApJ...859...47A}}). {\refv The NANOGrav Collaboration uses the Green Bank Telescope (GBT) at the Green Bank Observatory and {\reff formerly used} the {\refo 300-m William E.\ Gordon telescope at the Arecibo Observatory (AO).}}  

Rapidly rotating neutron stars, observed as pulsars, are remarkably stable astrophysical clocks. Acquiring a {\refb useful} TOA dataset for the purpose of GW detection involves collecting pulse arrival times every {\refb one to four} weeks from each pulsar in the PTA for a decade or more. In the case of NANOGrav {\refz (\citealt{alam20a}, \citealt{alam20b})}, $\sim$20\,min observations are taken every one to three weeks, depending on the pulsar. Pulses are averaged or ``folded'' {\refbfr at the appropriate pulse period}, from which we obtain an average pulse profile shape over the observation's duration. Each folded set of pulses is assigned an arrival time in frequency-dependent sub-bands. The folded pulse profiles are then {\refb compared to} a timing model that accounts for the many phenomena that influence arrival times. The list includes many timing perturbations from the ionized interstellar medium, or IISM, between Earth and {\refb the} pulsar (\citealt{2013CQGra..30v4006S}, \citealt{2018JPhCS.957a2007D}), such as variations in dispersion measure (DM; a line-of-sight integral of the electron density in the IISM), scattering, and {\refyy interstellar scintillation (ISS)}. {\refw These effects are all chromatic, unlike frequency-independent effects such as {\refxxx spin noise} intrinsic to the pulsar {\reff and any GW signals present during observations}.}

{\refo Significant} {\refz physical} processes not explicitly included in a timing model will lead to {\refo significantly} non-zero timing residuals, the differences between the measured arrival times and {\refs those predicted by} the timing model. {\refb After accounting for these astrophysical and systematic effects, the final residuals may reveal evidence for GWs, or provide upper limits on their magnitude} {\refz \citep{2013CQGra..30v4015S}}. {\refw Modeling chromatic timing effects also provides invaluable ancillary data for studying turbulence, lensing, and other plasma structures in the IISM (\citealt{Lam_2018}, \citealt{2019BAAS...51c.492S}).} 

Building on {\refz studies of} PSR~B1937+21 {\refq (\citealt{2011MNRAS.416.2821D}, \citealt{2013ApJ...779...99W})}, {\refz and of other simulations \citep{2015ApJ...815...89P},} this paper {\refyy explores the} conditions under which {\refb scattering can be {\refq deconvolved} using a signal processing technique known as cyclic spectroscopy (CS).} {\refb Section~2 describes some specific aspects of the IISM {\refq to which deconvolution can be applied}. Section~\ref{sec:mcyc} derives a figure of merit that can be used to predict how well coherent deconvolution will work for a given pulsar.} In Section~\ref{sec:simcyc} we describe simulated datasets that represent a variety of pulsar observations. Section~\ref{sec:quiltresults} shows the results of the simulations, {\refoo with comparisons to observations}. {
\refv Section~\ref{sec:benefits} makes predictions about the number of pulsars that could become PTA-quality after applying CS. We discuss implications for GW detection {\refxxx and for radio astronomy in general} in Section~\ref{sec:theend}.}
\vspace{-1em}

{\refb \section{The Transfer Function of the IISM}}
\label{sec:transfer}

{\reff To a high degree of accuracy, the ionized ISM acts as a linear filter on the radio waves passing through it \citep{han71}. Using standard terminology from signal processing, we can characterize this either by the  complex-valued voltage impulse response function (IRF), denoted $h(t)$ or, equivalently, through the Fourier transform of $h(t)$, the transfer function $H(\nu)$ \citep{hr75}.}

{\refo {\refxx {\refb The dominant source of frequency-dependent delay due to the IISM is from plasma dispersion.
This is time-dependent for a particular line-of-sight because of the combined motion of the pulsar, the observer, and the intervening medium.}
The observed DM at any given moment introduces a TOA delay $\propto \textrm{DM}/\nu^{2}$, {\refb where $\nu$ is the observed radio frequency}. {\refbfr The effects of DM variations on timing are typically represented by a single time-varying number, the broadband DM, although a more complete model would incorporate a frequency-dependent DM \citep{2016ApJ...817...16C}}. {\refw The transfer function of the IISM is dominated by cold plasma dispersion, which can be modeled as an all-pass quadratic chirp filter. In modern pulsar timing observations, the majority of this dispersive effect is removed by applying the inverse filter function to the received voltage data in real-time, using a fixed value of the DM. This process is referred to as coherent dedipsersion \citep{hr75}. Stochastic and systematic variations about the fixed DM are then incorporated into the timing model.}
\vspace{0.5em}
{\refb \subsection{Interstellar Scattering of Pulsar Signals}}

In addition to dispersion delay, the radio signal is 
scattered by inhomogeneities in the IISM.
Since the unscattered angular size of the pulsar is exceedingly small,  multi-path scattering results in a variety of interference effects that can be used to assess and potentially mitigate the effects of scattering delay.
%This results in an additional delay in the measured arrival time. 
The scattering is traditionally described by two regimes: diffractive and refractive {\refw \citep{1990ARA&A..28..561R} although the demarcation is not always sharp}. 
Diffractive interstellar scintillation, DISS, refers to the short timescale (minutes to hours) variation as the observer moves through the random diffraction pattern caused by the scattered ray bundle.
This stochastic pattern can be easily visualized by plotting the pulsar spectrum as a function of time, resulting in a ``dynamic spectrum.''
Although random in nature, this pattern has a characteristic width in frequency and time resulting in bright islands of power known as ``scintles" surrounded by regions of relatively weaker signal.

Larger scale inhomogeneities in the IISM can cause the ray bundle to expand or contract or, in extreme cases, break into multiple components that would represent multi-imaging events if the necessary angular resolution was available. 
These refractive interstellar scintillation (RISS) effects have a timescale of days to weeks, governed by the length of time it takes for the multi-path ray bundle to move transversely to the line-of-sight by its width, thereby guaranteeing a fresh (uncorrelated) signal path.
It has been conventional to model the IISM as a single phase-changing screen with
a Kolmogorov density structure {\refb \citep{2006ApJ...637..346C}}, and we take that approach here.
Increasingly, however, pulsar scintillation studies are yielding a different picture of the IISM: one in which any sight line crosses multiple relatively thin scattering screens with highly non-stationary statistics \citep{2019ApJ...870...82S}.}

{\reff The IRF of the IISM is changing with time because of motion of the line of sight through the medium. It is useful, therefore, to define a long-time average of the intensity IRF. We will refer to this as the pulse broadening function (PBF), denoted as $\overline{h_I (t)} = \langle |h(t)|^2 \rangle$, with the angle brackets denoting a time average.}
A Kolmogorov spectrum of electron density fluctuations in {\reff a thin screen results in a PBF that is}
close in functional form to a one-sided exponential, although in practice the kernel has a broader wing {\refb \citep{2010ApJ...717.1206C}}. 

The {\reff amplitude of the} impulse response function (IRF) fluctuates about the {\reff square root of the} mean PBF. The instantaneous IRF depends on the configuration of the IISM at a given time during an observation.} The centroid of an {\refv PBF corresponds to a pulsar's characteristic scattering timescale, or $\tau_{\rm s}$, {\refo a quantity often cited in the literature \citep{2016ApJ...818..166L}}. The corresponding scintle width, or diffractive bandwidth $\Delta\nu_d$ \citep{1990ARA&A..28..561R}, is:
\begin{equation}
    \Delta\nu_d = \frac{C_1}{2\pi{\tau_{\rm s}}}
    \label{eqn:diff}
\end{equation}

\noindent where $C_1$ is a dimensionless constant of order unity. }{\refb The effect of scattering is strongly frequency dependent: $\tau_{\rm s} \propto \nu^{-4.4}$. The {\refoo diffractive timescale, the timescale on which the scintillation pattern changes (also known as the scintillation timescale)}, scales as $\propto \nu^{-2.2}$ \citep{2010arXiv1010.3785C}}, and is typically minutes to hours for PTA observations. {\reff (The actual frequency dependencies for any given line-of-sight may differ significantly; our method in this paper does not rely on any particular exponent.)}

A first-order correction of scattering delays as a function of time can be accomplished by simply subtracting  $\tau_{\rm s}$ {\refoo from TOAs}. This reduces a scattering delay to a single number instead of an entire function. The $\tau_{\rm s}$ value is typically obtained through an autocorrelation function (ACF) analysis of the scintillation structure, given that $\tau_{\rm s}$ is inversely proportional to the scintillation bandwidth. {\citet{2016ApJ...818..166L}} measures the long-term $\tau_{\rm s}$ for NANOGrav pulsars, {\refo while \citet{turner20} provide} an updated analysis on the recent {\refo NANOGrav} 12.5-year data set. {\refb {\refxxx However, the correction by this single parameter $\tau_{\rm s}$ does not account for the distortion of the {\refm IRF shape away from the one-sided exponential ensemble average PBF}, which may lead to higher order timing errors. On the other hand, by determining the IRF as a function of time and frequency through the cyclic spectrum, we can remove both the delay and the distortion of the pulse profile by a deconvolution.}}

{Coherently deconvolving the IRF from the measured voltage waveform {\refz intrinsic to the} pulsar provides a more accurate correction by separating the intrinsic pulsar signal from the scattering transfer function of the IISM \citep{jonescyclic}. Several papers demonstrate a method for doing this using CS (\citealt{2011MNRAS.416.2821D}, \citealt{2013ApJ...779...99W}).  This technique relies on computing the cyclic spectrum of the voltage waveform, which is defined for cyclostationary signals---{\refq e.g.} any noise modulated by a periodic envelope (\citealt{2007MSSP...21..597A}, \citealt{1991ISPM....8...14G}). Pulsar emission {\refo is a particularly} good example of a cyclostationary signal.}

{\refw The {\refb deconvolution} algorithm discussed in the following sections will be referred to as WDS, in reference to \citet{2013ApJ...779...99W}.}
{\reff We note that the approach explored here and pioneered in early cyclic spectroscopy papers relies on a minimum of assumptions about the problem:  the pulse profile and IRF do not change over the time span analyzed; and each pulse consists of uncorrelated noise \citep{hr75}.
In principle, no assumptions are needed about the distribution of scattering material along the line of sight or the inhomogeneity spectrum of that material. 
This is in contrast to intensity-only techniques that have been employed in the past (e.g. \citealt{Bhat04}; \citealt{lohmer04}; \citealt{geyer16}; \citealt{geyer17}) that require specific assumptions about the functional form of the scattering kernel.}

%\newpage
%\vspace{1em}
{\refxx \subsection{The Signal Model}}
%%\vspace{1em}
%{\refb \subsection{The Effects of Interstellar Scattering on Observations}}

When a pulsed signal arrives {\refb at a telescope}, the standard practice is to {\refxxx detect pulses by dedispersing and folding data} in real time. {\refyy The full {\refo electric field (E-field)} information is typically discarded to keep the recorded data volume manageable.} Pulsar data therefore tends to be an average of pulse intensity profiles (for a particular polarization), $I(\nu,t) = |E(\nu,t)|^2$. {\refbfr For coherent {\refxxx WDS} deconvolution, the {\refq phase information must be preserved in a cyclic spectrum}. In practice, the receiver band is divided into frequency channels with a digital filterbank and the {\refo complex} voltage time series {\refo corresponding to} $E(\nu,t)$, for each frequency channel $\nu$, is recorded to disk.} Through the methods outlined here, the ultimate aim of deconvolution will be to understand how the E-field of the original {\refw pulse train} is changed by its interaction with the IISM (\citealt{2005MNRAS.362.1279W}, \citealt{2008MNRAS.388.1214W}). 

To describe the process of CS deconvolution, we take a simple model for {\refo the E-field of} a pulsar signal passing through the IISM and telescope:
{\refbfr \begin{equation}
{\refb E(t) = [p(t)N(t)] * h(t) + n_{\textrm{sys}}(t)},
\end{equation}}\noindent
where $p(t)$ is the original (unconvolved) pulse profile {\refw at time $t$ mod $P$}, $N(t)$ is the intrinsic modulated pulsar noise, $h(t)$ is the IRF,

{\refbfr $P$ is the pulse period,} and {\refb $n_{\textrm{sys}}(t)$} is the {\refw sky and receiver noise present in the system, uncorrelated across pulse periods}. 
This can also be written as
{\refbfr \begin{equation}
E(t) = X(t) * h(t) + n_{\textrm{sys}}(t),
\end{equation}}\noindent
in which $X(t) = p(t)N(t)$. In the frequency domain the signal model becomes:
\begin{eqnarray}
E(\nu) & = & [p(\nu) * N(\nu)]H(\nu) + n_{\textrm{sys}}(\nu) \\
& = & X(\nu)H(\nu) + n_{\textrm{sys}}(\nu),
\end{eqnarray}

\noindent{\refw using $X$ instead of $p * N$ because the convolution occurs upon emission at the pulsar}.

The \emph{cyclic spectrum} of $E(t)$ is:
\begin{equation}
%{\refb \begin{multline}
{S_E}(\nu, \alpha_k) = \langle E(\nu + \alpha_k/2)E^{*}(\nu - \alpha_k/2)\rangle,
%\end{multline}
\end{equation}

\noindent where $\nu$ is the radio frequency at when the signal is measured and $\alpha_k = k/P$ is the \emph{cyclic frequency}, also known as the \emph{modulation frequency}. {\refxxx The integer $k$ also refers to the number of pulse periods {\it in the time domain} that we would shift the signal and its conjugate before taking their product.} {\refw The mean here is taken over an integer number of pulses.} The {\refo resulting folded} cyclic spectrum {\refz is complex-valued} with amplitude and phase for each $(\nu, \alpha_k)$ pair, and is not defined for non-periodic signals.

\vspace{1em}
\includegraphics[width=.97\columnwidth]{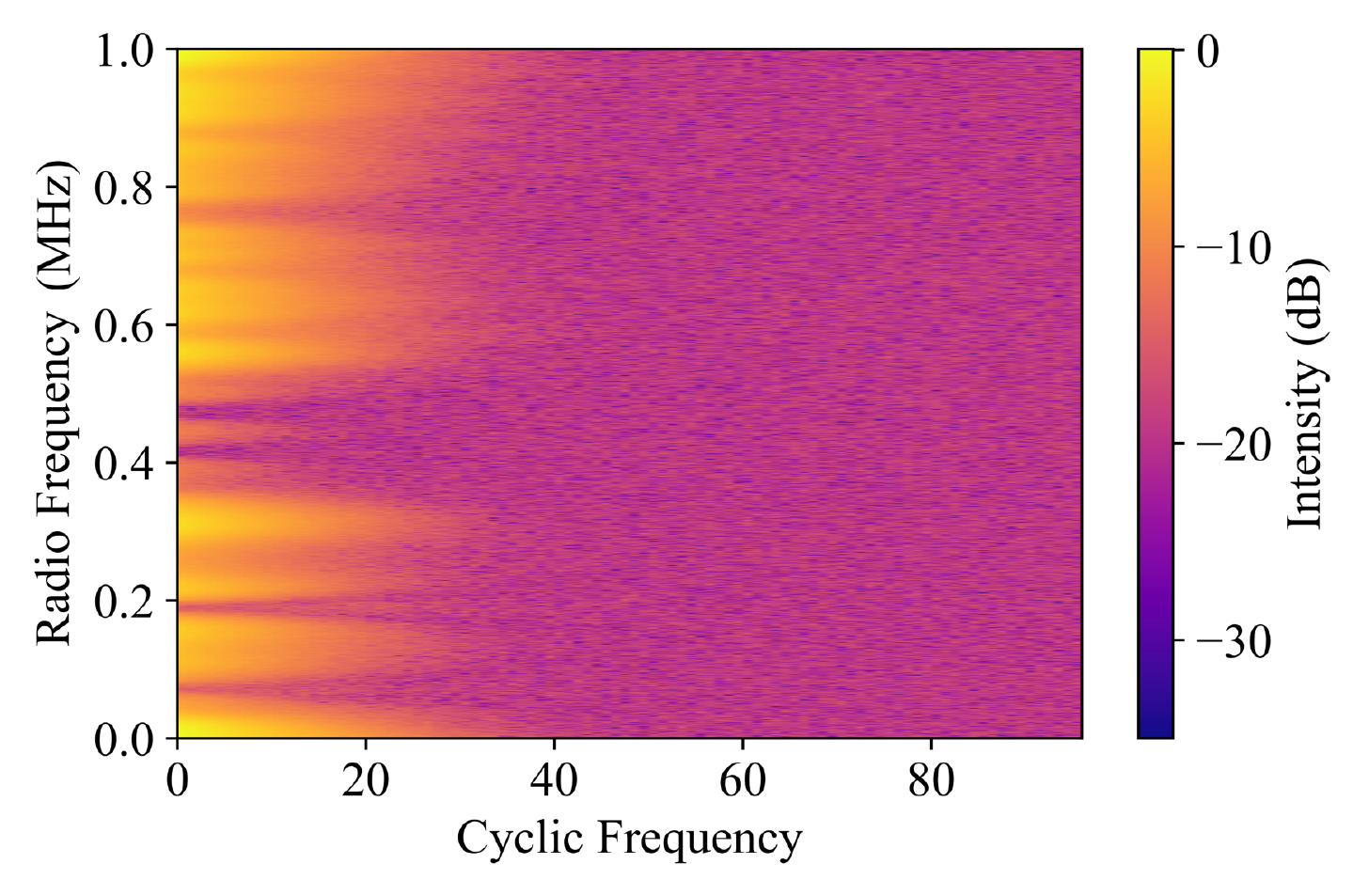}
%{cyc_amplitude.PNG}

\figcaption{Idealized, simulated example: {\refb magnitude of the cyclic spectrum for} a {\refx scattered} {\refxx pulsar} signal. {\refo The color scaling is logarithmic in power with an arbitrary reference level.} {Power is seen as a function of both cyclic frequency $\alpha_k$ {\refm (Hz)} and radio frequency $\nu$ {\refm (MHz)}, chosen in this simulated example to be {\refxx a 1~MHz band} near 435~MHz.}\label{fig:cycamp}}
%\end{minipage}

\vspace{1em}

Encapsulating the complete effect of the IISM on the intrinsic (pre-ISM) {\refb pulsed signal} requires preserving {\refb the} E-field phase information. If the IISM were not present, we would have $h(t) = 1$. For every short integration time on the order of the scintillation timescale {\refw we can construct a cyclic spectrum as a function of $\nu$ and $\alpha_k$ instead of an intensity spectrum as a function of $\nu$ only}. The cyclic spectrum {\refxxx can be rewritten in} the following form:
{\refb \begin{multline}
{S_E}(\nu, \alpha_k) = \langle{H(\nu + \alpha_k/2)H^{*}(\nu- \alpha_k/2)} \\ \times X(\nu + \alpha_k/2)X^{*}(\nu - \alpha_k/2)\rangle
\\ = \langle{H(\nu + \alpha_k/2)H^{*}(\nu- \alpha_k/2)}\rangle{S_x(\alpha_k)},
\label{eqn:cycspec}
\end{multline}}\noindent
where $S_x(\nu, \alpha_k)$ {\refxxx is the Fourier transform (FT) of the {\refoo folded} intrinsic pulse profile, which we will write as} $S_x(\alpha_k)$, assuming that there is negligible pulse profile evolution with radio frequency across the relevant observing band. {\refxxx Because of this assumption, we calculate cyclic spectra over fairly narrow bandwidths, meaning that variations in radio frequency in that spectra will be due to the IISM alone.} The {\refb shifts by $\pm\alpha_k/2$} represent the range of possible phase differences induced in the E-field from the IISM. Due to the periodic modulation (i.e. the intrinsically cyclostationary nature) of the intrinsic pulse, {\refxxx the modulus of $S_x(\alpha_k)$ is also {\refoo called} the harmonic profile.} {\refb The cyclic spectrum, ${S_E}(\nu, \alpha_k)$, is {\refxxx then} a 2D array that contains both {\refoo folded} E-field amplitude and E-field phase} information. {\refb ``Phase'' in this context specifically refers to the} electromagnetic phase induced only by the propagation of the intrinsic pulses through the IISM, {\refb not the pulse phase}. The FT of the cyclic spectrum along the $\alpha_k$-axis yields the cyclic periodogram, {\refxxx which shows the pulsar signal averaged over pulse phase and radio frequency}. The cyclic periodogram can produce finer frequency resolution than a standard filterbank for periodic signals. 

\vspace{1em}

\section{{\refxxx The Cyclic Merit, A Metric of Deconvolution Quality}}
\label{sec:mcyc}
Regardless of any later analysis one wishes to perform, the cyclic spectrum is itself an efficient way of collecting and storing the phase information from a given pulsar observation without resorting to large baseband datasets. {\refxx The cyclic spectrum is {\refo also} a tool for IISM deconvolution: the form of the spectrum expressed in Equation~\ref{eqn:cycspec} suggests that $H$ can be separated from $S_x(\alpha_k)$ in principle.} In practice this separation is non-trivial.

Detailed derivations of relevant CS quantities and their expected noise properties are provided in \citet{2011MNRAS.416.2821D} and \citet{2013ApJ...779...99W}. {\refb In order to predict the fidelity of the IRF determined by the WDS algorithm, we derive a figure of merit related to the signal-to-noise ratio (S/N) of the cyclic spectrum.} 

%\begin{minipage}{\columnwidth}
%###\vspace{1em}

%%\vspace{-4em}

%\end{minipage}
%###\vspace{-1em}
\newpage
{\refxx \subsection{A Description of the Cyclic Spectrum by Example}}

{\refz We first consider} some properties of the cyclic spectrum in the {\refx presence} of interstellar scattering. For pulse-modulated noise, we consider the cyclic spectrum of a scattered pulse. {\refxx In this example}, the cyclic spectrum corresponds to a {\refxxx 1.6\,ms {\refoo pulse period (the same as PSR~B1937+21)} with a $\sim$5\% duty cycle scattered with a 4-$\mu$s scattering tail.} {\refm These values represent a typical MSP that NANOGrav would routinely observe, with the period chosen to match that of PSR~B1937+21.} {\refoo The amplitude of {\refb the cyclic spectrum} is simply $\left|H(\nu)\right|\left|{S_x(\alpha_k)}\right|$ (see Figure \ref{fig:cycamp}). } The examples shown {\refb} in Figures \ref{fig:cycamp} and \ref{fig:cycphase} have {\refx a pulse profile S/N of 70.} {\refoo In the blue regions of Figure~\ref{fig:cycphase}, {\refw the phase of the cyclic spectrum $\Phi_{S_E}$ remains coherent} with $\alpha_k$}. {\refb Deconvolving IISM effects using CS hinges on this {\refw phase {\refoo structure}}, which represents the degeneracy between $H(\nu+\alpha_k/2)$ and $H(\nu-\alpha_k/2)$ being broken.} In contrast, {\refbfr a non-periodic source {\refb such as a quasar (in this noiseless case)} would have a zero-valued cyclic spectrum} in amplitude everywhere except for the $\alpha_k = 0$ column, which is the standard spectrum of the source (Figure~\ref{fig:cyccollapsed}), and would have an entirely incoherent {\refx complex phase across radio frequency}. 
\vspace{1em}
\includegraphics[width=.97\columnwidth]{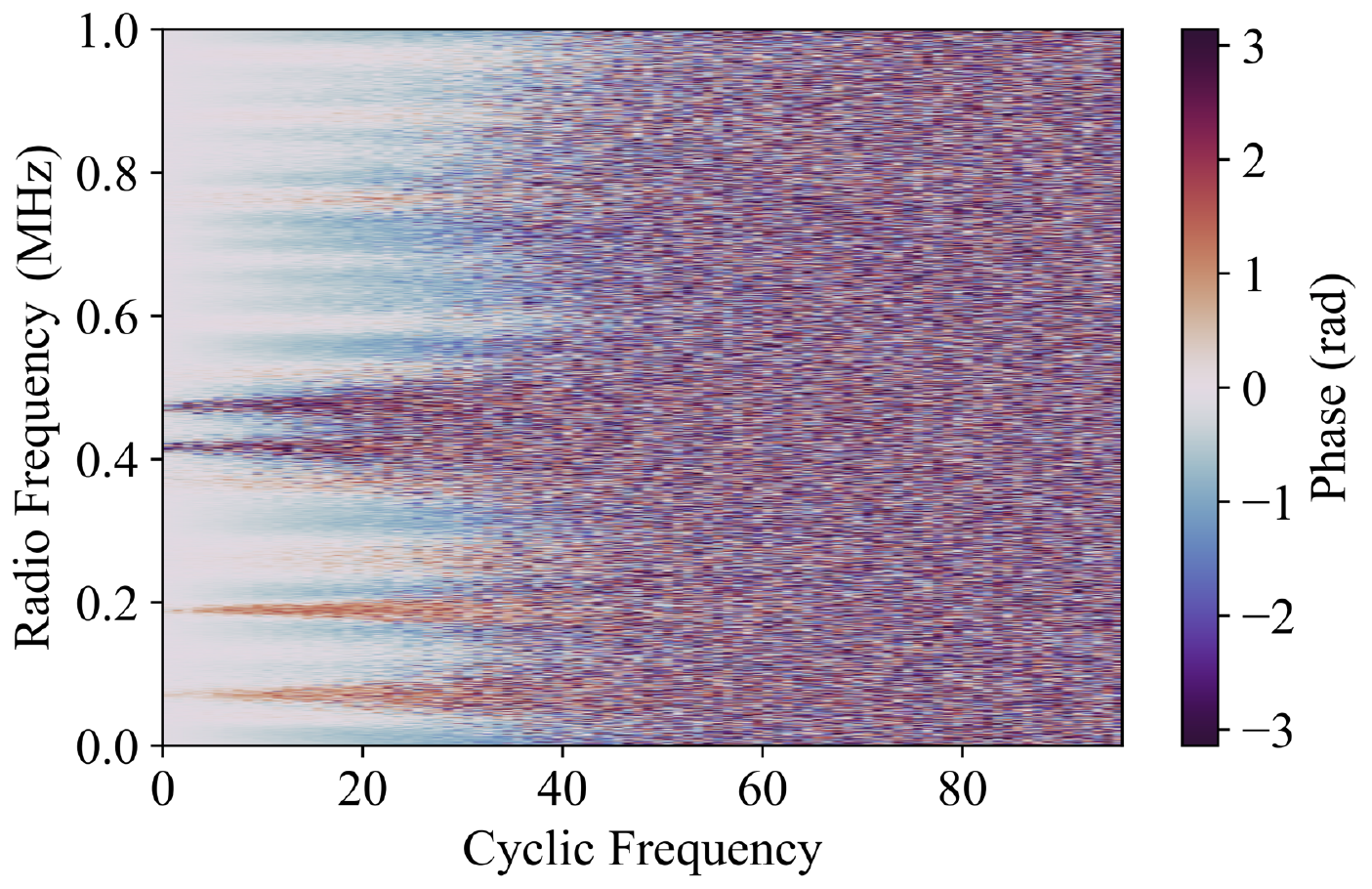}
\figcaption{Idealized, simulated example: {\refb phase of the cyclic spectrum for} a {\refx scattered} {\refxx pulsar} signal. The {\refv color scaling is $\pm\pi$ in phase}. Phase is seen as a function of both cyclic frequency $\alpha_k$ {\refm (Hz)} and radio frequency $\nu$ {\refm (MHz)}, chosen in this simulated example to be over 1~MHz near 435~MHz. {\refv The zero-to-negative phase transition within the scintles is the phase slope resulting from the IISM. The cyclic spectrum makes this phase slope apparent. {\refoo Figure~\ref{fig:cycamp} is scaled logarithmically; by contrast, the phase structure in this figure, scaled linearly, goes out to much higher harmonics than does the amplitude.}}\label{fig:cycphase}}

%###\vspace{-2em}
\vspace{1em}
{\refq \subsection{A Figure of Merit for Deconvolution by Cyclic Spectroscopy\label{sec:figofmerit}}}

The WDS algorithm searches for both the true (unscattered) pulse profile {\refb and for the IRF that best explains the measured CS}. The algorithm performs this search by taking an initial guess at the IRF, usually a delta function, which corresponds to the specification {\refb $\partial{H(\nu)}/\partial\nu = 0$} (no scintles) and a constant phase. {\refb The $M$ complex samples that make up the IRF are then varied as a $2M$-dimensional parameter space}, holding the assumed intrinsic profile constant. While this methodology inherently introduces many possible degeneracies, \citet{2013ApJ...779...99W} show that the cyclic spectrum structure due to scattering will not be significantly covariant with the intrinsic profile. {\refb The intrinsic profile can then be determined by iteratively solving several consecutive cyclic spectra, while refining the estimate of the intrinsic profile with each subsequent iteration {\refxxx (the ``outer loop'')}.} The assumed intrinsic profile (at any step of the outer loop that iterates through {\refb successive cyclic spectra}) and the IRF that yielded the best-fit cyclic spectrum represent {\refb estimates of the intrinsic pulse profile and the IRF of the IISM}. The WDS procedure {\refbfr in \citet{2013ApJ...779...99W}} yields an IRF for the original millisecond pulsar B1937+21 {\refbfr at 430~MHz}, {\refxx in which} the best fit IRF {\refbfr clearly has an even more finely resolved scattering structure than that shown by earlier methods \citep{2001ApJ...549L..97S} in which AU-sized scattering structures in the IISM were observed.} {\refb The intrinsic profile is consistent between two {\refw nearby} frequencies treated independently, showing a high degree of convergence.}
%%%%\vspace{1em}

\vspace{0.15em}

\begin{center}
%\vspace{-6em}
\resizebox{1.\columnwidth}{!}{\includegraphics{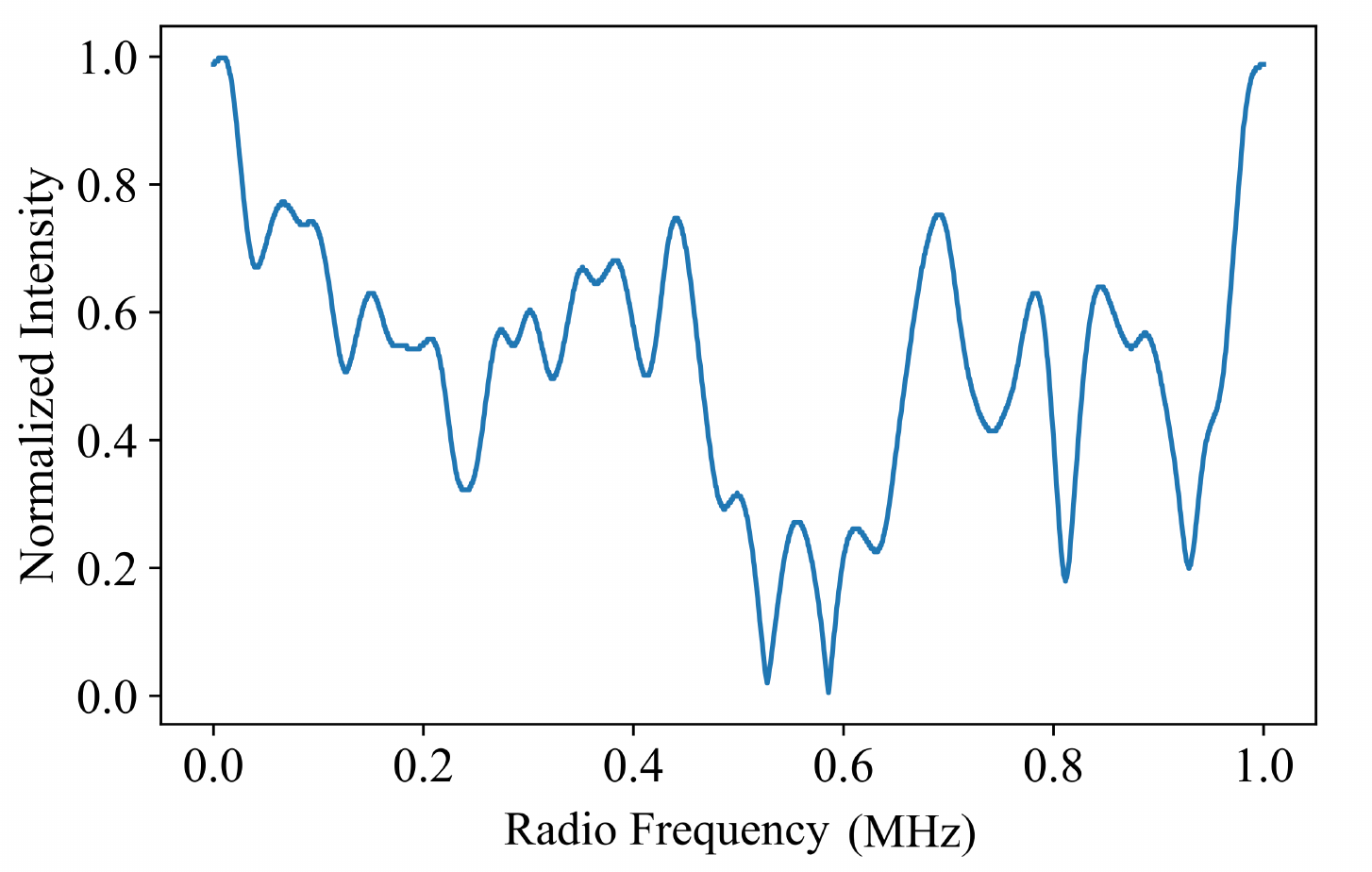}}
%\resizebox{\columnwidth}{!}{\includegraphics{fig3.pdf}}
%\vspace{-8em}
\figcaption{{\refx The radio frequency spectrum for the same simulated cyclic spectrum shown in Figures 1 and 2. This is the zero-valued column of Figure~1 along the {\refo radio} frequency axis. Each peak is a ``scintle'' due to ISS.} \label{fig:cyccollapsed}}

\end{center}

{\refb In our implementation of the WDS algorithm, we do not use the outer loop. This is because each simulation is limited to a single cyclic spectrum made from a single realization of an IRF. One could simulate a realistic sequence of IRFs that evolve with time to more fully characterize the WDS algorithm, but doing so is beyond the scope of this paper. {\refoo In real observations, once a best-fit intrinsic profile is found, it can be used in repeated epochs, as it will generally vary much less significantly than the IRF on a diffractive timescale, likely making the inner loop more critical.}}

{\refb We find that} WDS fits for the cyclic spectrum amplitude $\left|{{S_E}(\nu,\alpha_k})\right|$ in a relatively small number of iterations. {\refb This is not surprising because $\left|{S_E}(\nu,0)\right|$ is proportional to $\left|H(\nu)\right|$, {\refoo providing} a good initial estimate of the magnitude.} As we move out to higher $\alpha_k$ in $\left|{{S_E}(\nu,\alpha_k)}\right|$, the functional form across {\refo radio} frequency is narrowed and overtaken by noise. {\refb Fitting the phase of ${S_E}(\nu,\alpha_k)$ requires considerably more effort by the non-linear optimization at the heart of WDS, for the following important reason.} {\refb Radio frequency channels between scintles have little to no signal at low values of $\alpha_k$, which results in regions of phase ambiguity. Qualitatively observing the evolution {\refz of the} fitting process {\refw in our runs of the code}, the complex phase of $H(\nu)$ often appears to converge long after its complex amplitude.} See \citet{2013ApJ...779...99W} for a more detailed explanation.
%%\begin{center}
%%\resizebox{\columnwidth}{!}{\includegraphics{complexampsumharm.pdf}}
%\resizebox{\columnwidth}{!}{\includegraphics{fig4.pdf}}
%%\figcaption{ \figfourcap }
%%\end{center}

The transfer function phase due to interfering E-field phases from the cyclic spectrum can be obtained as follows. In order to deconvolve the IISM along every ray path, we need E-field phase information contained in {\refbfr the phase of the transfer} function $H$ of the data we are fitting. Taking the complex phase of Equation~6, we have: %The best-fit transfer function comes from the output from WDS.
\begin{equation}
\Phi_{S_E}(\nu, \alpha_k) = \Phi_H(\nu + \alpha_k/2) - \Phi_H(\nu - \alpha_k/2) + \phi_{S_x}(\alpha_k)
\end{equation}
\vspace{-1em}
\begin{equation}
= \alpha_k\frac{\Phi_H(\nu + \alpha_k/2) - \Phi_H(\nu - \alpha_k/2)}{\alpha_k} + \phi_{S_x}(\alpha_k)\end{equation}
\noindent where $\Phi_E$ is the phase of the cyclic spectrum at $\nu$ and $\alpha_k$, $\Phi_H$ is the phase of the transfer function $H(\nu,\alpha_k)$, and $\phi_{S_x}$ is the phase of the intrinsic profile harmonics. In the limit {\refo $\alpha_k \ll \Delta\nu$, {\refoo where $\Delta\nu$ is the radio frequency channelization,}} we have:
\begin{equation}
\phi_{S_E}(\nu, \alpha_k) \approx \alpha_k\frac{\mathrm{d}\Phi_H}{\mathrm{d}\nu} + \phi_{S_x}(\alpha_k)
\label{eqn:phaseslope}
\end{equation}
\vspace{-1em}
\begin{equation}
= \frac{k}{P}\frac{\mathrm{d}\Phi_H}{\mathrm{d}{\nu}} + \phi_{S_x}(\alpha_k).
\label{eqn:phaseslope2}
\end{equation}
\noindent
\noindent
{\refv The problem then becomes solving} for $\Phi_H$ and $\phi_{S_x}$ in Equation~\ref{eqn:phaseslope2}. The WDS algorithm {\refo arrives at} a solution through optimization techniques. The likelihood of extracting useful phase information from the cyclic spectrum can be found by calculating a figure of merit, which is the expected average value of $\Phi$ divided by the uncertainty in that estimate (see Appendix A for derivation and details):
%\begin{equation}
%\label{eqn:cycmerit}
%\end{equation}
\begin{equation}
    m_{\rm cyc} = 
    \frac{\Phi}{\delta\Phi} =2\pi\frac{\tau_s W_e}{P^2} (S/N) \sqrt{ \sum_k k^2 a_k},
    \label{eqn:cycmerit}
\end{equation}
where $\tau_s$ is the scattering time, $P$ is the pulsar period, $W_e$ is the equivalent width
\footnote{ $W_e$ is the width of a top-hat pulse with the same maximum value as a
folded profile \citep{2004hpa..book.....L}.}
of the average pulse profile,
$S/N$ is the averaged signal-to-noise ratio {\refo} of the maximum of the pulse profile divided by the off-pulse\footnote{ CS does not require specifying the arbitrary ``on'' and ``off'' pulse regions used in standard radio pulsar analysis. Instead, the $\alpha_k$ structure contains generalized information about the strength of the repeated signal across the profile. {\reff We assume, however, that the pulse profile has been smoothed to the optimal sharpness width, $W_s$, defined in the Appendix.}}, and $a_k\equiv A_k/A_0$, {\refoo where the $A_k$ values are the amplitudes of the Fourier transform of the intensity pulse} profile {\refu (note the importance of CS for MSPs, with $m_{\rm cyc} \propto P^{-1}$, {\refm assuming $W_e \propto P$})}.

For $m_{\rm cyc} \gg 1$, {\refb the cyclic spectrum should provide enough information for WDS to successfully deconvolve {\refoo the IRF}} to a reasonably high accuracy. Testing this assumption with simulations is critical because it is not obvious how specific features of the algorithm might limit {\refbfr recovery}, such as {\refz phase-wrapping}, in which the phase of $H(\nu)$ might be lost between scintles. Many other possible subtleties of WDS described in {\refb \citet{2013ApJ...779...99W}} could also {\refb reduce the effectiveness of the algorithm.}

{\refw A high S/N ratio is clearly a criterion for successful deconvolution.} In one sense, a long scattering tail {\refb provides more phase slope {\refv (see Figure~\ref{fig:cycphase})} that can be leveraged by the deconvolution algorithm {\refxxx if it rises above the phase noise as we move toward higher harmonics or $\alpha_k$ values}.} However, a highly scattered pulsar may still be more difficult for the WDS algorithm to deconvolve, given the phase-wrapping considerations just mentioned. {\refb {\refoo We now proceed to} describe tests of the effectiveness of the WDS algorithm on simulated datasets and evaluate the results in relation to the cyclic figure of merit.} 
%\begin{center}
%\includegraphics{scienceseminar5.pdf}
%%%\vspace{3em}
%\end{center}
\begin{figure*}
\begin{center}
%\vspace{-10em}
\resizebox{2.1\columnwidth}{!}{\includegraphics{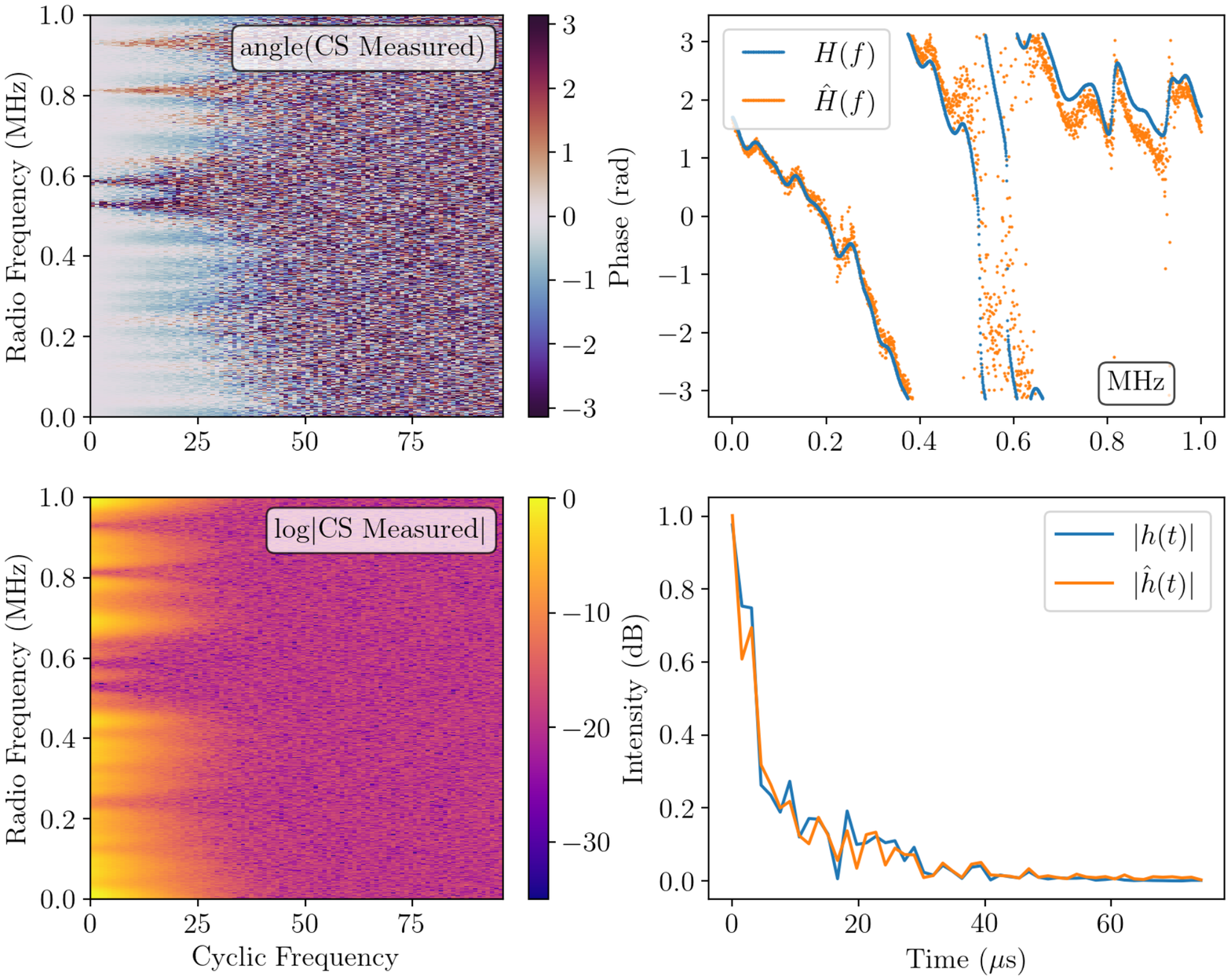}}
\vspace{-2em}
\figcaption{Simulated recovery of an impulse response function (IRF) for an artificial pulsar signal with period {\refb 1.6\,ms} and scattering {\refb time constant} $\tau_{\rm s}$ = 4\,$\us$, $S/N = 70$. {\refbfr The top left plot shows the phase of the cyclic spectrum {\refm as a function of cyclic frequency in Hz}, including a {\refw stable phase} due to scattering, similar to the toy model in Figure \ref{fig:cycphase}, with the color scaling representing complex phase. The bottom left plot shows the amplitude of the cyclic spectrum {\refm as a function of cyclic frequency in Hz}, similar to the toy model in Figure~\ref{fig:cycamp}, with a color scaling logarithmic in power. The upper {\refv right figure shows the complex phases of the simulated (blue) and recovered (orange) transfer functions, or FTs of the IRFs. {\refn The x-axis shows frequency (in MHz).} The bottom right plot shows the simulated (blue) and recovered (orange) impulse responses in the time domain, for only the first 75 of 2048 samples, {\refxxx scaled so that the peak value is 1 dB}. See Section~\ref{sec:mcyc}}} for {\refbfr further} details. {\refn The plots are generated from the \textsc{pycyc} code.} \label{fig:simexample}}
\end{center}
\end{figure*}
\vspace{0.5em}
\section{{\refxxx Deconvolving} The Simulated Datasets}
\label{sec:simcyc}
{\refxxx {\refoo We generated a suite of artificial IRFs and recovered them using WDS, looking at these results by varying the S/N of the de-scattered pulse profile and the simulated $\tau_{\rm s}$ in order to determine the region in {\refxxx the $\tau_{\rm s}$-$S/N$} parameter space in which CS is effective. WDS was re-implemented in Python in the publicly available \textsc{pycyc} code\footnote{\url{https://github.com/gitj/pycyc}}}}.
%\vspace{-0.5em}
\subsection{Setup and Properties of Simulation Suite}\label{subsec:setup}
We assumed 2048 frequency bins over a bandwidth of 1\,MHz, {\refxxx in order to ensure resolved scintles; thousands of channels of frequency resolution across several scintles is routinely possible after employing CS, as with the periodic spectrum of PSR~B1937+21 in \citet{2011MNRAS.416.2821D}. The very narrow radio frequency channelization assumed here is not strictly necessary for the success of CS deconvolution; what is necessary is a large enough {\refoo $\mathrm{d}\Phi_H/\mathrm{d}{\nu}$ to accumulate CS phase while avoiding obscuration by random variations in $\phi_{S_x}(\alpha_k)$} as in Section~\ref{sec:figofmerit}. The simulated pulsar has a duty cycle {\reft for its intrinsic pulse shape} of about 5\%, similar to PSR~B1937+21 \citep{1998ApJ...501..270K}.}

{\refbfr We generated transfer {\refxxx functions each with characteristic width $\tau_{\rm s}$ by using the relevant routines in \textsc{pycyc} to multiply the exponential IRF envelope by complex Gaussian noise. We then added noise directly to the simulated cyclic spectrum to simulate the presence of} radiometer noise. In both stages the noise was white and Gaussian in each of the real and imaginary components. The one-sided exponential with multiplicative noise becomes the IRF which the WDS algorithm will recover.} {\refxxx  An entirely realistic pulse profile would include two components of white noise in the time domain, one additive, corresponding to radiometer noise, and one component of amplitude-modulated noise (AMN), emanating from the emission mechanism from the pulsar itself. Doing so is beyond the scope of this paper, as such an arrangement would be too expensive computationally.} Adding artificial noise directly to the cyclic spectrum is slightly unrealistic because each element of the 2D cyclic spectrum array receives an independently generated noise value. In reality, the noise and signal values in a cyclic spectrum are not independent from bin to bin because {\refb the values result from the correlation products of the measured voltage data}. The purpose of the simulation, however, is to explore the conditions under which a phase slope can rise above the noise present in a cyclic spectrum, regardless of the {\refb detailed} characteristics of that noise.

Figure \ref{fig:simexample} shows an example simulation of a realistic IRF {\refw and its resulting scintillation structure {\refoo after a converged WDS fitting process}}. The bottom right panel shows {\refb the time domain magnitude of} the simulated IRF (blue) and the recovered IRF ({\refo orange}). The signal bandwidth is 1\,MHz and the pulse period is 1.6\,ms in the artificial pulsar signal. {\refb The number of frequency channels is chosen to reduce computation time. It is not necessary to simulate a range of bandwidths because the important quantity to vary is the number of scintles across the band.} {\refxxx In the example shown in the figure, {\refbfr $\tau_{\rm s}$ is the $1/e$ timescale in }the blue curve in the bottom right.} For a 1.6\,ms pulsar, 5\,$\mu$s of scattering is a small fraction of the pulsar's {\refb period}, {\refxxx which is why the {\refo noisy} one-sided exponential is only visible in the first $\sim$20 of the 2048 time bins of the IRF}. The upper {\refo right} panel shows the complex phase of the transfer function, or FT of the IRF. {\refb As demonstrated in Section~2}, the slope of the {\refw phase of the {\refo transfer function}} is proportional to the {\refw scattering timescale}, or the centroid of the IRF in the bottom right panel. The bottom left and upper left panels show, respectively, the amplitude and phase of the complex cyclic spectrum array, to which WDS was applied and from which the results in the upper right and lower right panels were extracted. The color scale of the amplitude plot {\refbfr is in} logarithmic units, and the scaling of the phase plot ranges from $-\pi$ to $\pi$. In the amplitude of the cyclic spectrum, {\refb the magnitude of $H(\nu)$ is discernible along the vertical direction}. Each scintle decays with increasing harmonic in the horizontal direction. In the phase of the cyclic spectrum a {\refxxx slope} is apparent in the horizontal direction. The degree to which this phase stability {\refbfr is measurable} {\refxxx across harmonics} becomes a diagnostic for how well IISM deconvolution is possible{\refb .}

{\reff The cyclic merit $m_{\rm cyc}$ of the example in Figure \ref{fig:simexample} is $\sim$4. While the quality of recovery in this case is reasonably good by eye, the low cyclic merit value implies that this example is an outlier amongst iterations that would generally not allow for reliable CS deconvolution.} 

{\refb The upper right panel of Figure \ref{fig:simexample} also demonstrates the inherent difficulty in the WDS fitting process caused by phase ambiguities.} {\reff In the example shown, the lowest quality fit around 0.6\,MHz is caused by a low S/N in that region of the cyclic spectrum, which is visible in the upper left panel.} For a pulsar signal scattered {\reff significantly more than this example}, the number of phase wraps may become so large that the phase of $H(\nu)$ may become indistinguishable from noise {\refv in the low-amplitude regions of the bandwidth}.% {\refb 

{\refxxx Our goal is to find the quality of IRF reconstruction in both complex amplitude and phase. We define 36 ``cells’’ in a two-parameter grid of simulated pulsar signals, varying pulse profile S/N and scattering timescale parameters, to see how the WDS algorithm behaves in practice for a variety of possible astrophysical signals.}

We chose {\refb scattering times of 1, 2, 4, 8, 16, 32, 64, 128, and 256\,$\mu$s} for the $\tau_{\rm s}$ of a B1937-like pulsar (period 1.6\,ms). {\refo {\refb We used} {\refxxx full-bandwidth intrinsic} pulse profile S/N {\refb (peak to {\refoo off-pulse rms})} values of} 20, 70, 650, and 2600. {\refb For each combination of $\tau_{\rm s}$ and $S/N$, we ran at least 60 simulations} with different random number seeds determining both the radiometer noise and the realization of the scattering variations modulating the one-sided exponential. Each simulation had {\refb a limit of 3000 objective function evaluations by the non-linear optimizer}. {\refbfr Convergence occurs when successive function values differ by a factor of} {\refb approximately} the machine $\epsilon$, {\refbfr or the smallest numerical step possible on a given processor,} which is an extremely high standard of convergence. {\refbfr Most runs converged long before 3000 {\refb objective function evaluations}.}

\subsection{Measuring the Quality of Impulse Response Function Recovery}
\label{subsec:calib}

{\refoo Once WDS determined a best-fit IRF for a particular simulation, we calculated the} quality of recovery by comparing the input and output IRFs\footnote{In {\refxxx reality, the true complex IRF will have an amplitude of $\sqrt{2/\tau_{\rm s}}$ guaranteeing its normalization to 1, but here were are only interested in recovering the functional form of the complex IRF. Once a best fit IRF and (via an outer loop) intrinsic profile are obtained, their convolution can always be re-scaled to match the flux of the measured profile.}}. For the $>$60 simulations in a {\refbfr cell}, {\refv we define a goodness-of-fit metric $\Delta$ as:
\begin{multline}
    \Delta \equiv \frac{1}{3\tau_{\rm s}}\sum^J_{i=0}\biggl(\frac{h_\textrm{input}(t_i)}{\sum^J_{j=0} h_\textrm{input}(t_j)} \hspace{1em}- \\  \frac{h_\textrm{output}(t_i)}{\sum^J_{j=0} h_\textrm{output}(t_j)}\biggr)^2
\end{multline} 
\\
\noindent {\refo where $J$ is determined such that $J\delta{t} = 3 \tau_s$, where $\delta{t}$ is the sampling interval ($t_J = 3\tau_{\rm s}$)}.
By construction, the simulated IRFs ({\refoo each $h_\textrm{input}$} above), are one sided exponential functions with $1/e$ widths of $\tau_{\rm s}$ and multiplicative random noise but otherwise noiseless. The multiplicative noise is a simulated astrophysical signal with each IRF representing a different realization of the IISM. The radiometer noise, the observational noise added to the simulated cyclic spectra, is present in both the recovered intrinsic pulse profile and the recovered IRF. Normalizing IRFs to a value of 1 is difficult in the presence of the additive noise acquired in the WDS fitting process, in which each iteration of $h$ depends on most recent best-fit cyclic spectrum, because the normalization will be influenced by the noise present in $h(t)$ when $t \gg \tau_{\rm s}$. We instead perform our normalization only out to $3\tau_{\rm s}$, where the IRF signal is present, and then take the sum-of-squares value out to $3\tau_{\rm s}$ as well. We finally divide by $3\tau_{\rm s}$ to compare the goodness-of-fit per time bin, so as be able to compare simulations with different $\tau_{\rm s}$ values. The parameter $\Delta$ is taken to be the typical quality of reconstruction for a given $\tau_{\rm s}$ and S/N ratio. {\reff In the deconvolution example of Figure~\ref{fig:simexample}, $\logten{\Delta} = -3.1$, which is $\sim$1.3$\sigma$ above the mean of our simulated $\logten{\Delta}$ values (lower $\logten{\Delta}$ representing better fit quality), consistent with the $m_{\rm cyc}$ range that represents varying recovery quality as stated in \ref{subsec:setup}.} 

{\refbfr  {\refxxx {\refoo The quantity $\Delta$ used in this work is different than the ``demerit'' used in \citet{2013ApJ...779...99W}, a quantity which compares the data to a model cyclic spectrum at each step of} the fitting process. The WDS algorithm minimizes the demerit. Here, we are instead comparing a simulated IRF to the recovered best-fit IRF.}} In Appendix B of \citet{2013ApJ...779...99W}, the authors show that quality-of-fit (the positive curvatures or second derivatives of demerit) is proportional to $F^2$, where $F$ is the total pulsed flux in a pulsar signal. For a fixed S/N in our simulations, where S/N is the peak to {\refoo off-pulse rms}, increasing $\tau_{\rm s}$ is equivalent to increasing $F$ in the suite of simulated pulsars. For our high $\tau_{\rm s}$-valued and/or high S/N-valued individual simulations, then, the metric $\Delta$ should be closely related to the best-fit demerits from the WDS algorithm. However, the $\tau_{\rm s}$ and $S/N$ values at which the recovered IRFs become unreliable is not straightforward, which is why we propose $m_{\rm cyc}$ as a new metric, to be verified in the following section with simulations.
}
\begin{figure*}

%\begin{center}
%\vspace{-10em}

%\vspace{-5em}
%\hspace{.4in}
\vspace{-8.5em}
\hspace{-.1in}
\resizebox{2.4\columnwidth}{!}{\includegraphics{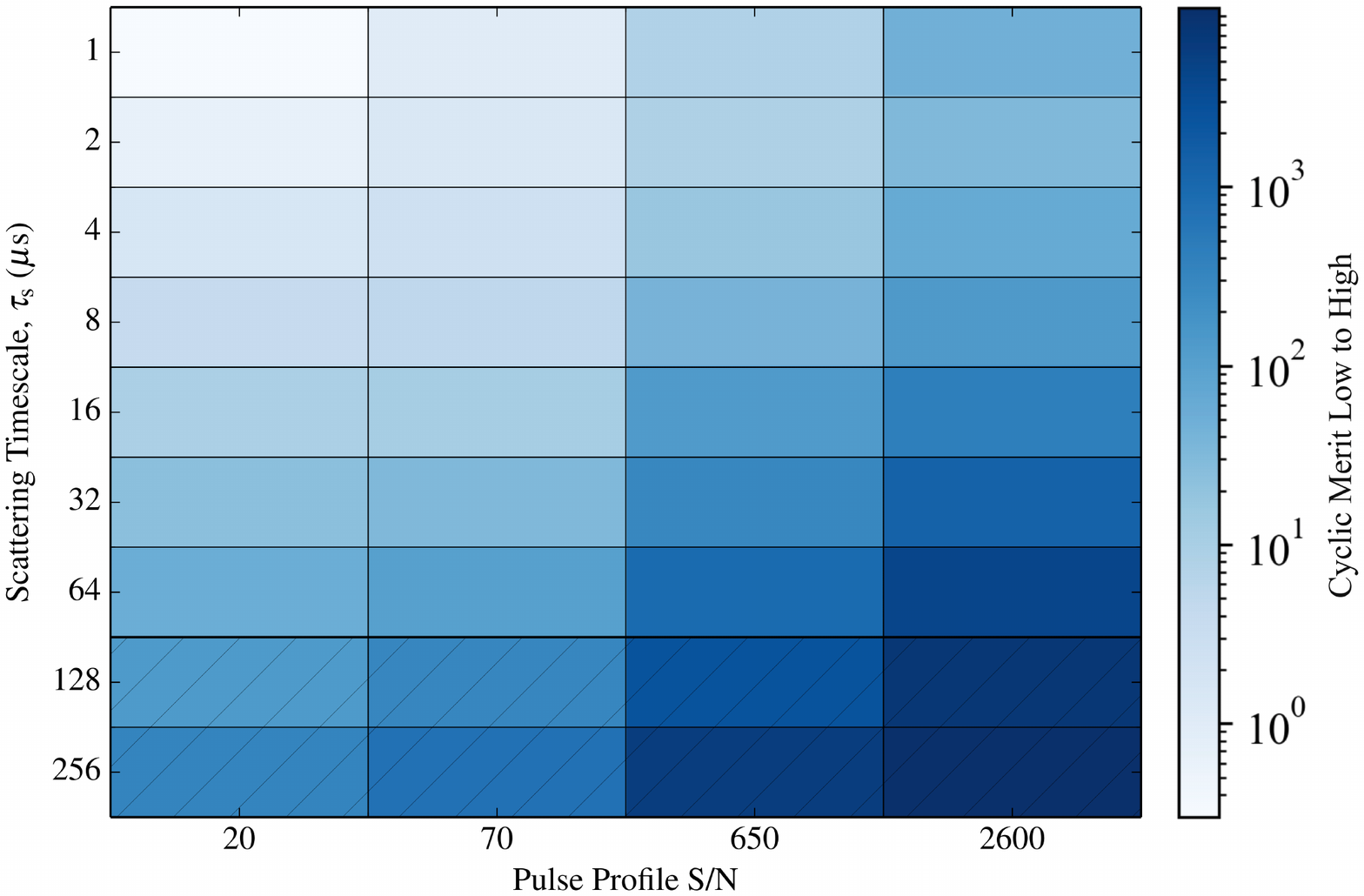}}
%\vspace{-14em}
\vspace{-8em}
\figcaption{Quality of impulse response function recovery for {\refv simulated pulsars} using the WDS algorithm (Section~\ref{subsec:simmerit}). {\refv Higher values in colorbar mean better quality of recovery. {The values in each cell were originally goodness-of-fit metrics $\Delta$, then reassigned to cyclic merit values $m_{\rm cyc}$ (Equation \ref{eqn:cycmerit}), as described in Section~5.2. Values of $\tau_{\rm s}$ are in $\mu$s. Each cell corresponds to $>$60 simulations, with the color corresponding to the median $m_{\rm cyc}$ value. Simulations in the crosshatched region are those in which scintles are not resolved by the fixed number of radio frequency channels across the simulations performed here, in which case the recovered IRFs are broad one-sided exponentials. {\refn The simulations verify that the effectiveness of CS deconvolution is proportional to both pulse profile S/N and scattering timescale $\tau_{\rm s}$.} }\label{fig:quilt}}}

%\end{center}
\end{figure*}

%\newpage
\section{Results and Comparisons with Observations}
\label{sec:quiltresults}

{\refn Using the simulations, quality measurements, and theoretical cyclic merit from the previous two sections, we compare a suite of simulations to test the dependence of $m_{\rm cyc}$ on $\tau_{\rm s}$ and $S/N$. We also compare several observations to the theoretical cyclic merit predictions, establishing a fiducial metric for the reliability of routine deconvolution of the IISM.}
%\vspace{-1em}
\subsection{Measuring the Simulated Cyclic Merit Relationship}
\label{subsec:simmerit}

{\refb Results from the simulations are summarized in Figure \ref{fig:quilt}. Each {\refv cell in the grid} corresponds to a combination of $\tau_{\rm s}$ and S/N ratio.} The  color scale corresponds to the logarithm of the $m_{\textrm{cyc}}$ values described in Section~\ref{sec:mcyc}. {\refv The $\Delta$ values from each corresponding simulated and recovered IRF pair were calibrated to the theoretical $m_{\textrm{cyc}}$ values by the procedure to be described in this section, and then rescaled to arrive at the merit-based color scaling in Figure \ref{fig:quilt}. Note that while the vertical {\refoo axis shows $\tau_{\rm s}$ values that are powers of two}, the $S/N$ values in the horizontal axis increase by {\refo a factor of} $\sim$3.5 except from 70 to 2600 which increases by almost two of the logarithmic intervals. {\refo We did not simulate the missing middle column to save computing time}; the gap is taken into account in {\refo our linear regression procedure}.
}

For the B1937+21-like artificial pulsar used in the simulations {\refv(similar in period and in main pulse width, with no interpulse in simulations), we relate our goodness-of-fit metric $\Delta$ to $m_{\rm cyc}$ by:
\begin{equation}
\Delta = {A}m_{\textrm{cyc}}^{-\mu}
\label{eqn:powerlaw}
\end{equation}
\noindent because we expect an inversely monotonic relationship between $\Delta$ and $m_{\textrm{cyc}}$ across many orders of magnitude in $\tau_{\rm s}$ and $S/N$, but we do not know the details of how the two will correspond at low $m_{\textrm{cyc}}$, as discussed in {\refoo Section~\ref{subsec:calib}}. We assume $\mu$ is a constant with respect to $\tau_{\rm s}$ and $S/N$. Although we already {\refo have a theoretical expectation for} how $m_{\textrm{cyc}}$ will depend on $\tau_{\rm s}$ and $S/N$ from Equation \ref{eqn:mcycsn}, we parameterize possible powerlaw dependencies as:
\begin{equation}
m_{\textrm{cyc}} \propto \tau_{\rm s}^{\beta}\left({\mathrm{S/N}}\right)^{\gamma}.\label{eqn:reg}
\end{equation}
We are interested in verifying that the simulation results demonstrate that $\beta = 1$ and $\gamma =1$, which would be consistent with the derivation in Appendix A. Re-writing this in logarithmic form, Equation~\ref{eqn:reg} becomes: 
\begin{equation}
\logten{m_{\textrm{cyc}}} = \beta{\logten{\tau_{\rm s}}} + \gamma\logten{\left(\mathrm{S/N}\right)} + C
\label{eqn:firstlog}
\end{equation} \noindent
where $C$ is a constant derived from pulsar-specific parameters in Equation \ref{eqn:mcycsn}. We can then substitute Equation \ref{eqn:powerlaw} into Equation \ref{eqn:firstlog} to obtain: 
\begin{equation}
-\frac{1}{\mu}\logten{\Delta} = \beta{\logten{\tau_{\rm s}}} + \gamma\logten{\left(\mathrm{S/N}\right)} + D
\end{equation}
\noindent where $D$ depends on both $A$ and $C$. Proceeding, we note that because:
\begin{equation}
\logten{\Delta} = -\mu\beta{\logten{\tau_{\rm s}}} -\mu\gamma\logten{\left(\mathrm{S/N}\right)} - \mu{D}
\end{equation}
\noindent we cannot measure $\beta$ and $\gamma$ from the results of our simulations because each will be covariant with $\mu$. However, we can measure the ratio $\gamma/\beta$, which should be 1, by least-squares linear regression. This is equivalent {\refoo to measuring the direction of the gradient of $\logten{\Delta}$ but not its magnitude}. Because we have derived $\beta = 1$ and $\gamma =1$ in Appendix A, the gradient of $\Delta$ should point in the $(1,1)$ direction in $\logten\tau$-$\logten(\mathrm{S/N})$ space. We ignore the constant $\mu{D}$ in our fitting.

The distribution of the $\Delta$ values within each cell was approximately Gaussian. Across cells, the median $\Delta$ values appear approximately coplanar in Figure \ref{fig:quilt} by eye. An exception is the upper right cell which is a local minimum compared to the surrounding cell. We examined the individual input and output IRFs and found that because the S/N was very high in this region, most $\Delta$ values were very near zero. The statistical distribution of the $\Delta$ values was different from {\refoo the apparently Gaussian distribution} in other cells, {\refo most likely} because the statistics are determined more by the numerical properties of the computational process than by the simulated IISM properties. In the following fit, we neglect this outlier upper-right cell.

\begin{figure*}
%\begin{center}
\vspace{-8.25em}
\hspace{-.1in}
%\hspace{-10em}
\resizebox{2.4\columnwidth}{!}{\includegraphics{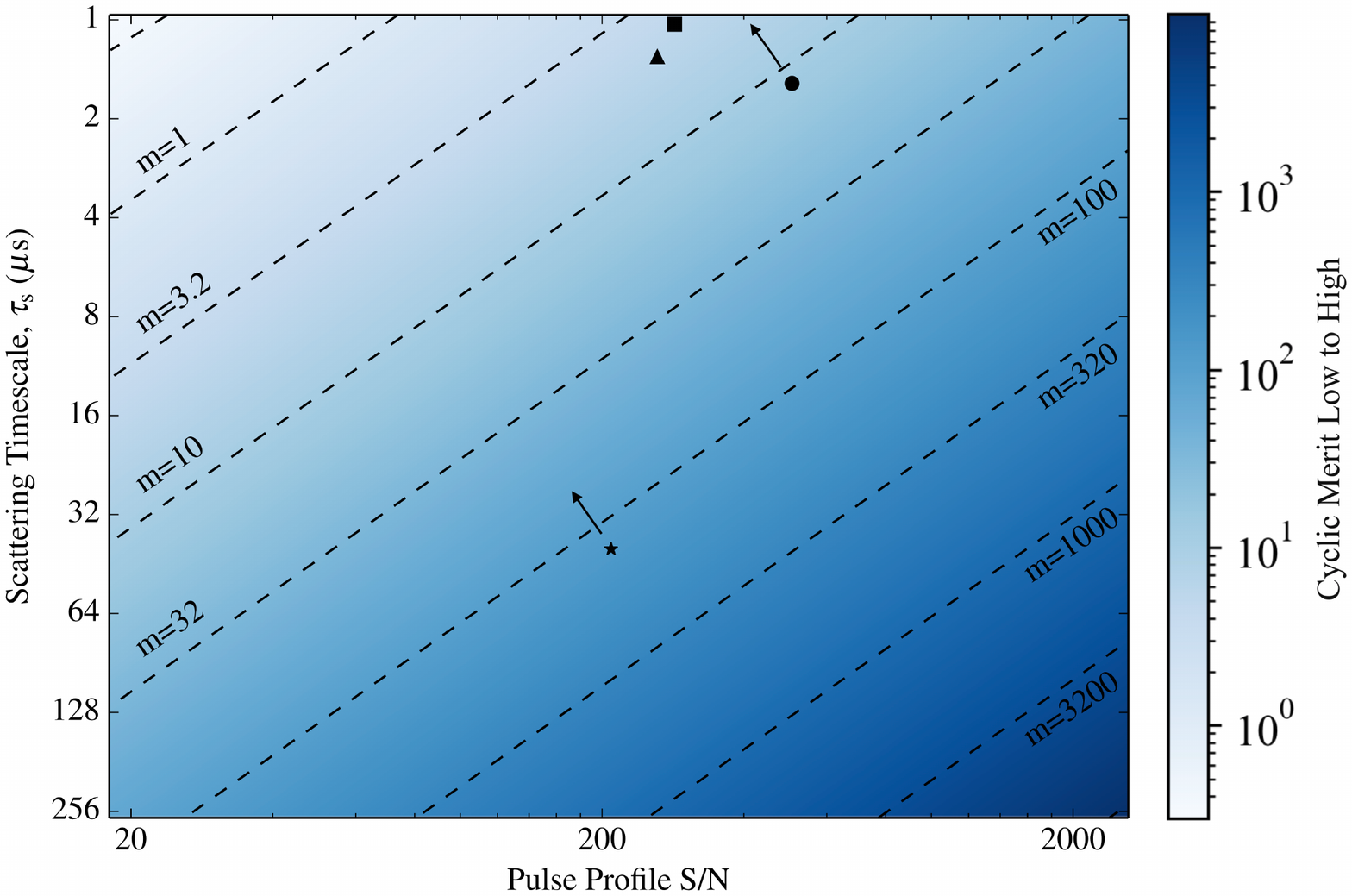}
}

%\vspace{-14em}
\vspace{-8em}
\figcaption{{\refxx {\refn Theoretical quality of impulse response function recovery (Section~\ref{subsec:comparemerit}) from Equation~\ref{eqn:cycmerit}. Four examples of CS deconvolution on real pulsars are also shown:} PSR~J2317+1439 at 327~MHz (triangle), PSR~B1937+21 at 430~MHz (star; from \citealt{2013ApJ...779...99W}), PSR~J1713+0747 at 327~MHz (square), and PSR~B1937+21 at 1410~MHz (circle).} Of the four real pulsar examples, the upper three real pulsars did not successfully recover an IRF, which is consistent with their low cyclic merit ($m_{\rm cyc}$) values. {\refn The predicted cyclic merit values shown {\refo in the colorscale} are for a generic pulsar with the period and equivalent width of PSR~B1937+21, and the pulse profile shape of the PSR~B1937+21 main pulse only. Pulsars J1713+0747 and J2317+1439 therefore have $m_{\rm cyc}$ values that differ slightly from those shown. The arrows next to both observations of PSR~B1937+21 indicate that in reality, the interpulse reduces the cyclic merit by a factor of $\sim$2, resulting in three of the four observations with $m_{\rm cyc} < 10$, those three which we also found to not regularly converge on a best-fit IRF. Note that a pulsar's S/N is entirely {\refoo dependent on the telescope sensitivity and on the folding time}. Upcoming telescopes will move more pulsars to the right. Inclusion of significantly scattered pulsars not currently used in timing programs will further populate the bottom of the plot with high $m_{\rm cyc}$ pulsars.\label{fig:quiltdata}}}
%\end{center}
\end{figure*}

A two-parameter unweighted linear regression of $\logten{\Delta}$ with $\logten{\tau_{\rm s}}$ and $\logten{\mathrm{S/N}}$ yields slopes of $-\mu\beta = -1.86\pm0.07$ and $-\mu\gamma = -1.92\pm0.07$ respectively, with an $r^2$ regression statistic of 0.98. The slope ratio $\gamma/\beta$ is therefore $0.98\pm0.05$. With these nearly coplanar values and the expected gradient direction, we compute the $m_{\rm cyc}$ values from Equation \ref{eqn:cycmerit} for a B1937+21-like pulsar and assign those values to the cells, shown in the color scale of Figure $\ref{fig:quilt}$.

The bottom crosshatched two rows of Figure \ref{fig:quilt} are those in which the scintles are not resolved, and in which, as a result, the success of the WDS algorithm in those cells is not representative of the full capability of the algorithm. 

The radio frequency channel size here is 1~MHz$/2048 = 0.488$~kHz, which is held constant over all the simulations so that $\tau_{\rm s}$ and $S/N$ are the only variables changed. By Equation~\ref{eqn:diff}, for $\tau_{\rm s} = 64\us$, $\Delta\nu_d = 2.5$~kHz, and we have $\sim$5 channels per scintle. At $\tau = 128\us$ we have 1 or 2 channels per scintle, making resolving the scintles difficult. Because the scintillation structure across radio frequency corresponds to the multiplicative noise {\refoo structure} in an IRF, unresolved scintles are equivalent to a one-sided exponential IRF with no additional structure. {\refo The situation of unresolved scintles often occurs when observing pulsars at radio} frequencies $<100$~MHz; scattering timescales are measured by taking an intrinsic pulse shape derived from higher frequencies and convolving with one-sided exponentials until the model profiles match the data \citep{Bansal_2019}. The crosshatched rows represent {\refs simulations with no IISM information gained over the method of successive convolutions with smooth, one-sided exponential IRFs}. However, we are not aware of any reasons why WDS would perform poorly for the highly scattered pulsars represented in the bottom two rows of Figure \ref{fig:quilt} {\refo if we had used} a greater number of channels. The slope ratio $\gamma/\beta$ for Figure~\ref{fig:quilt}, removing both the upper right outlier and the crosshatched cells, is $1.02\pm0.08$ {\refoo with an $r^2$ regression statistic of 0.97}, consistent with the prediction from Equation~\ref{eqn:cycmerit}. {\refm The S/N in Figure~\ref{fig:quilt} is integrated over the band, not the single-channel S/N.}}
\vspace{-1em}
\subsection{Comparisons of Figure-of-merit with Observations}
\label{subsec:comparemerit}
{\refn In Figure~\ref{fig:quiltdata}, we show $m_{\rm cyc}$ predictions {\refoo using, as in Figure~\ref{fig:quilt}, the period and main-pulse $W_e$ for PSR~B1937+21 as representative quantities for a generic MSP. We then compare to data.}}

The observations used here, in addition to the 430\,MHz {data on PSR~B1937+21} used in \citet{2011MNRAS.416.2821D} and \citet{2013ApJ...779...99W}, were taken {\refo from} three separate campaigns at Arecibo on pulsars PSR~J1713+0747 (4.6\,ms period) at 327\,MHz, PSR~B1937+21 (1.6\,ms period) at 1.4\,GHz, and PSR~J2317+1439 (3.4\,ms period) at 327\,MHz. All three datasets were baseband. The PSR~J1713+0747 data (AO P2627, PI Palliyaguru) were taken at 430\,MHz on 19-Sep-2011 with Mock spectrometers as backend receivers and were 40\,MHz wide in bandwidth, {\refb with a 10\,MHz slice used for the deconvolution attempt}. The PSR~B1937+21 data (AO P2676, PI Dolch) were taken on 19-Sep-2012 and the PSR~J2317+1439 data (AO P2824, PI Stinebring) were taken on 02-Aug-2013, both with the PUPPI backend (Puerto Rican Ultimate Pulsar Processing Instrument; \citealt{2008SPIE.7019E..1AD}) with {\refoo a 200\,MHz and a 50\,MHz bandwidth}, respectively. {\refb Data were folded into cyclic spectra using the ``--cyclic'' option in the \textsc{dspsr} package {\refq (\citealt{2011PASA...28....1V}, \citealt{2011MNRAS.416.2821D})}. The deconvolutions used 25\,MHz of the available bandwidth {\refoo unless stated otherwise}, which simplified the dataset because of band gaps between channels. The cyclic spectrum was accumulated in integrations such that the duration of each integration was less than the diffractive timescale, so that the IRF would be stable throughout each integration.} {\refb The S/N of each data set was high enough that scintles were visually apparent in dynamic spectra formed from the data.} {\refv For most {\refb subintegrations} of data, the WDS procedure was applied and no converging IRF was found. We attempted a number of different starting conditions, varying the shift in phase between input profile and the measured profile, but in {\refb none of these cases} did the algorithm yield a coherent solution. {\refb The same stopping and convergence criteria were used {\refoo on both the data and} the simulations.} We found one well-fitting IRF in a subintegration of the PSR B1937+21 data at 1.5~GHz, but based on the cyclic merit values in Figure \ref{fig:quilt}, we expect an occasional convergence by chance that may nor may not represent the true underlying state of the IISM during that subintegration.} 

{\refb The examples of deconvolution attempts on real data} are plotted in Figure \ref{fig:quiltdata} along with the converging example from \citep{2011MNRAS.416.2821D} and \citep{2013ApJ...779...99W}. {\refbfr The $S/N$ values for the data new to this paper {\refb were computed using a 25\,MHz bandwidth}.} {\refb The three non-converging examples are in a region of $\tau_{\rm s}$-$S/N$ space predicted to have poorer deconvolution quality than the converging example by {\refo an order} of magnitude.} {The $\tau_{\rm s}$ values {\reft and each pulsar's measured frequency powerlaw used here are from the NANOGrav 12.5-year dataset \citep{turner20},  unless otherwise stated. {\refm We recognize these values are representative of highly variable IISM structures along a pulsar's line-of-sight (the variability motivating CS deconvolution) but many reported values in the literature differ. The $a_k$ values (Equation~\ref{eqn:cycmerit}) used were also from the NANOGrav 12.5-year dataset \citep{alam20a}.}}} {\refu {\refo The $m_{\rm cyc}$ value for PSR~J1713+0747 at 430\,MHz ($\tau = 1.0\us$; $S/N$ = 285) was 4.3. For PSR~B1937+21 at 1.4\,GHz\footnote{{\refn Here we use the $\tau_{\rm s}$ from the one well-fitting IRF.}} ($\tau_{\rm s} = 1.6\us$; S/N = 506), $m_{\rm cyc} = 12.2$. PSR~B1937+21 at 430\,MHz had $m_{\rm cyc} = 125$ ($\tau_{\rm s} = 40\us$; $S/N$ = 209, \citealt{2013ApJ...779...99W}, \citealt{2011MNRAS.416.2821D}, using the $-3.66$ frequency powerlaw dependence of $\tau_{\rm s}$ in \citealt{ram06} at a frequency of 1273\,MHz where our baseband observations were located). Finally, PSR~J2317+1439 at 327\,MHz ($\tau_{\rm s} = 1.3\us$; $S/N$ = 262) had $m_{\rm cyc} = 5.1$}.  

In reality, these four pulsar observations will correspond to different colorscales in Figure \ref{fig:quiltdata} because they have different values of $P$, $W$, and $\sum_k k^2 a_k$ in Equation \ref{eqn:cycmerit}, whereas the colorscale of Figure \ref{fig:quilt} is based on PSR~B1937+21. {The $m_{\rm cyc}$ values for PSR~J1713+0747 and PSR~J2317+1439 are slightly higher and lower respectively due to the sharper pulse of the second producing more $a_k$ terms in Equation~\ref{eqn:cycmerit}.} {{\refs {\refn Our stated $m_{\rm cyc}$ values for PSR B1937+21 are greater than the true $m_{\rm cyc}$ values by a factor of $\sim$2 because the presence of the interpulse reduces $\sum_k k^2 a_k$ in Equation~\ref{eqn:cycmerit}. The arrows in Figure~\ref{fig:quiltdata} {\refo point to} the true $m_{\rm cyc}$ contour line where the two PSR~B1937+21 observations should reside. For PSR~B1937+21 at 430\,MHz \citep{2013ApJ...779...99W}, the actual $m_{\rm cyc}$ value is 66.}
Because our simulations were done on a single-component pulse profile representing a generic MSP, in Figure~\ref{fig:quiltdata} we still plot the $m_{\rm cyc}$ value from PSR B1937+21's main component $m_{\rm cyc}$ only. The results remain consistent with {\refn the concept of} $m_{\rm cyc}$ as a unitless detection ``signal'' of a scattering tail. {\refn Taking the interpulse into account, the three unsuccessful CS deconvolution attempts have $m_{\rm cyc}<10$ and the successful attempt has $m_{\rm cyc} = 66$, which is why we use $m_{\rm cyc} = 10$ as an absolute minimum requirement for CS-enhanced pulsar timing in Section~\ref{sec:benefits}.}}} } 

The predictions in Figure~\ref{fig:quiltdata}, verified by the simulations in Figure~\ref{fig:quilt}, suggest that despite the {\refz large amount of phase-wrapping}, a pulsar in the bottom-left of Figure \ref{fig:quiltdata} -- highly scattered but low S/N -- could be successfully IISM-deconvolved. The large amount of scattering would mean a {\refbfr high} duty cycle in the scattered pulse profile. The de-scattered profile in such a case would both be scattering-corrected in terms of its TOAs {\refn (by eliminating variations in the IRF)}, as well as turned into a better precision-timed pulsar with its {\refbfr low-duty-cycle} unscattered pulse profile. {\refbfr However, this possibility would need case-by-case {\refm evaluation: as we move toward lower frequencies, $\tau_s$ increases, but the diffractive timescale decreases \citep{2010arXiv1010.3785C}, limiting the maximum integration time. A smaller $\Delta\nu_d$} also lowers the S/N per scintle.} A correct application of the figure-of-merit (Figure \ref{fig:quiltdata}) to pulsar observations should take these considerations into account.

A more detailed simulation will be necessary to evaluate the following cases, which depend on the pulsar and the observation in question:
%\vspace{-1em}
\begin{itemize}
\item If the pulsar flux becomes comparable to the telescope's system equivalent flux-density, then the on-pulse noise (due to the pulsar's own AMN) will significantly exceed the off-pulse noise, unlike our simulations which assume the same noise properties across all phases of the profile.  
%\vspace{-2em}
\item If the number of pulses averaged to obtain a cyclic spectrum is too low, then the effects of pulse phase jitter will be important \citep{Lam_2019}, and the correlations between ${S_E}(\nu,\alpha_k)$ bins in the cyclic spectrum array will become significant. Pulse phase jitter - the intrinsic white-noise distributed TOAs of single pulses - cannot be reduced by telescope sensitivity, as the TOA uncertainties are intrinsic to the pulsar. The TOA uncertainties can be reduced by longer integration times, but this is not helpful for mitigating scattering in real time, with the longest integration time on the order of a diffractive timescale. {\reft For highly scattered pulsars, the diffractive timescale may in some cases become so low that the necessary S/N for CS deconvolution is not possible\footnote{NANOGrav Memo \#4:~\url{http://nanograv.org/assets/files/memos/NANOGrav-Memo-004.pdf}}}.
\end{itemize}
As long as a particular pulsar observation is free of the above two properties, Figure \ref{fig:quiltdata} can be expected to reasonably predict whether CS deconvolution is feasible.%\newpage

{\refn Figure~\ref{fig:quiltdata} does not imply that CS is totally inapplicable to pulsar observations with $m_{\rm cyc} < 10$. Even if deconvolution is infeasible, a cyclic spectrum still {\refo contains a detailed measurement of} the presence of scattering, in addition to providing an extremely fine frequency resolution. In some cases, CS deconvolution may be possible even when $m_{\rm cyc} < 10$. What we have tested here is a straightforward, untailored application of the WDS algorithm. In practice, well-measured long-term PBFs for particular pulsars could serve as ideal initial conditions, instead of the delta functions used here. An ideal frequency binning could also change the behavior of the algorithm. Such approaches would model a line-of-sight and repeatedly update that model, much in the same way that pulsar timing models are repeatedly updated. The cyclic merit concept, then, must be understood as a measure of how reliably the WDS algorithm can quickly and routinely deconvolve, as in a real-time data processing pipeline.} 
\begin{figure*}
\begin{center}
\resizebox{1.3\columnwidth}{!}{\includegraphics{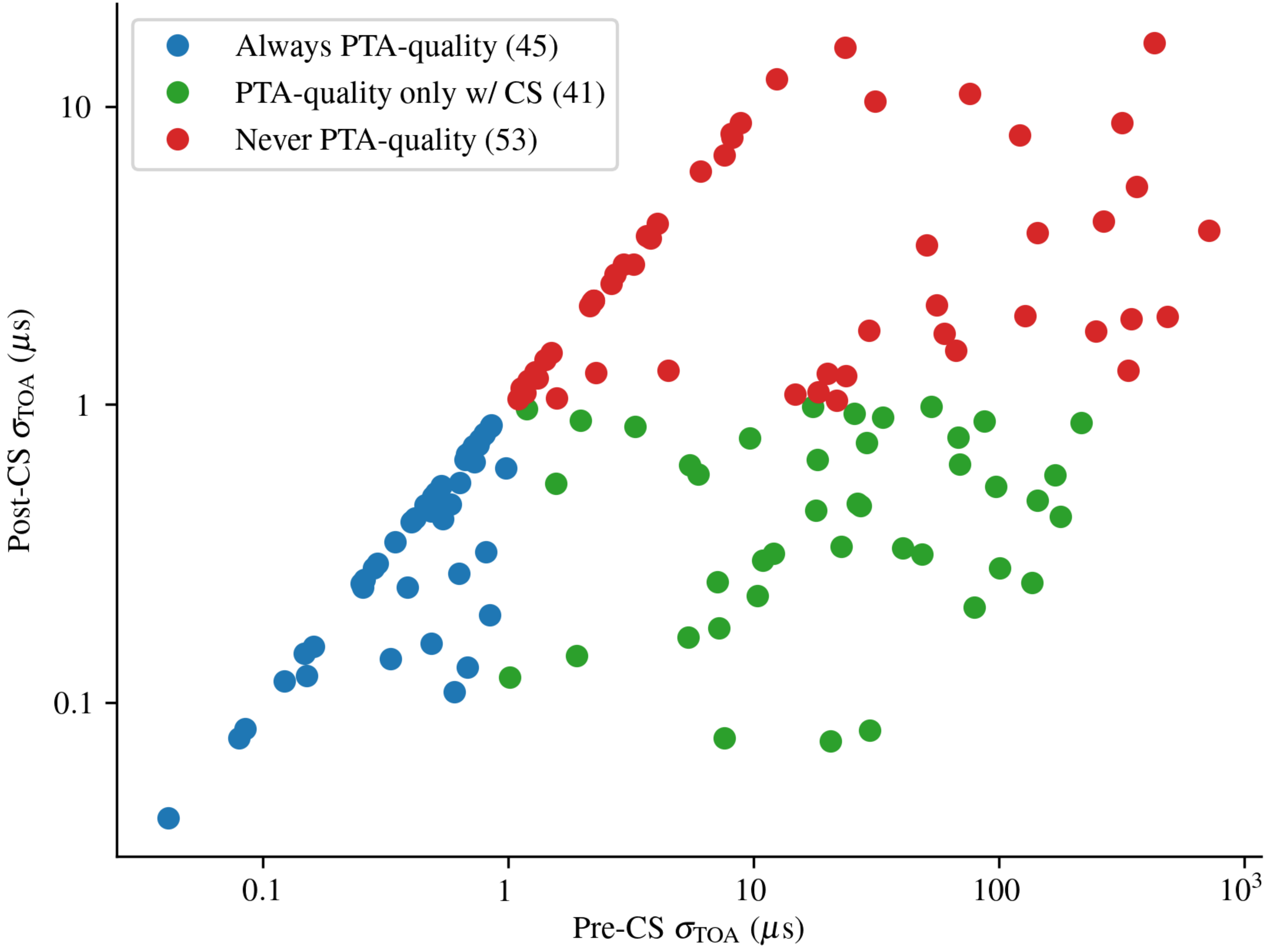}}

\figcaption{{\refv The improvement in timing precision
    for a simulated Galactic MSP population from the \texttt{PsrPopPy} code observed with the Green Bank Telescope's {\refo ultrawide bandwidth (UWB) receiver, currently under construction}.  Cyclic spectroscopy deconvolution has the potential to double the number of PTA-quality MSPs with
    $\sigma_{\rm TOA} < 1\; \us$ at the GBT using the {\refo UWB receiver}. The (blue) ``Always PTA-quality'' MSPs {\refm have similar properties to} current NANOGrav pulsars. The (green) ``PTA-quality only w/CS'' are those MSPs that could be included in the NANOGrav PTA because of the improved UWB sensitivity.} \label{fig:sims}}
\end{center}
\end{figure*}
%\vspace{-15em}
\begin{figure*}
\begin{center}
\resizebox{1.6\columnwidth}{!}{
\includegraphics[width=\columnwidth]{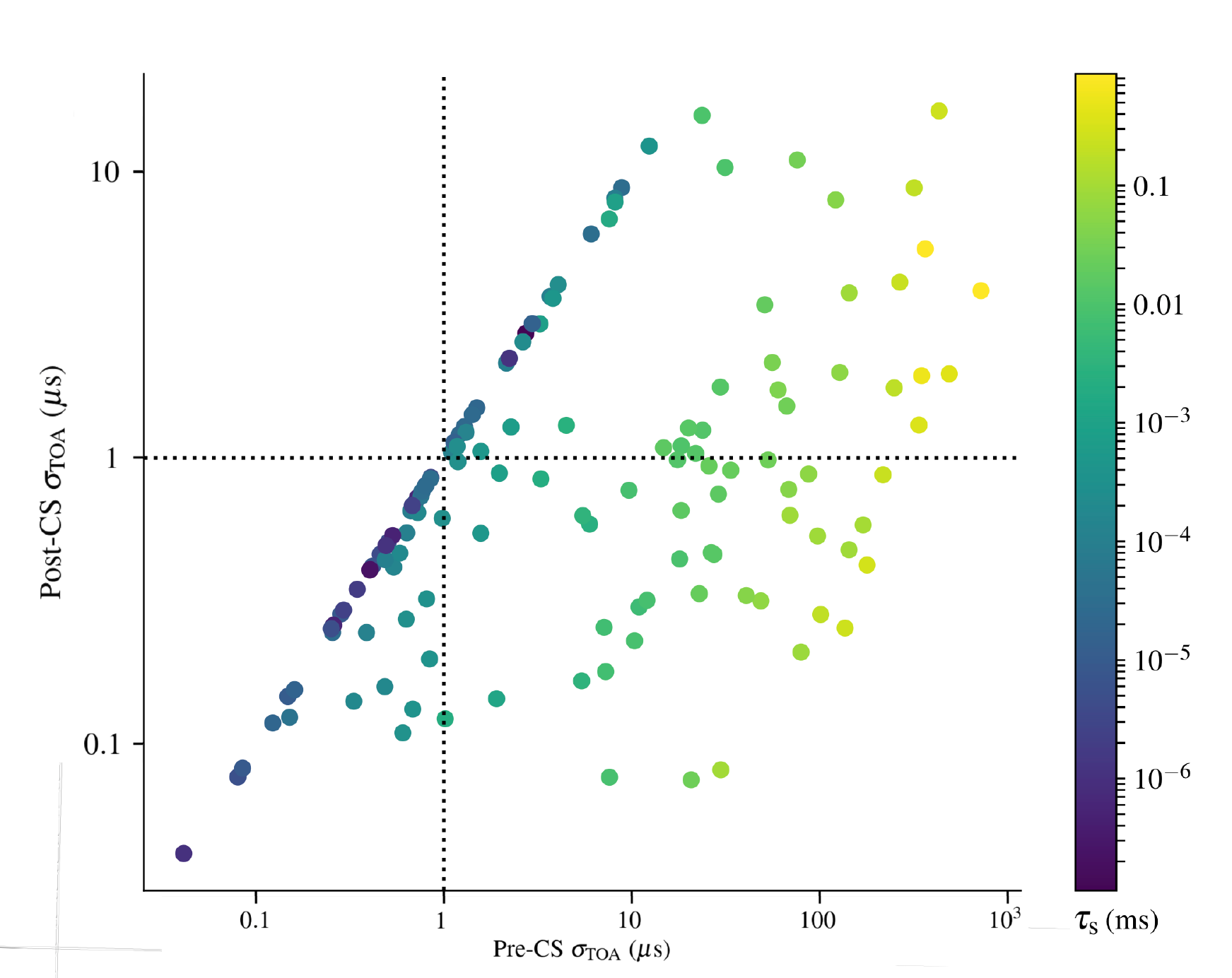}
}
\figcaption{{\reft The same simulated pulsars are plotted as in Figure~\ref{fig:sims}, but with the colorscale representing the $\tau_{\rm s}$ value of each simulated pulsar. The pulsars with the greatest timing improvements tend to be those with {\refo the longest} scattering tails, but the majority of the ``PTA-quality only w/CS'' pulsars have scattering tails with $\tau_{\rm s}$ values of $\sim10-100\us$, in accordance with Figure~\ref{fig:quilt}. The dotted lines correspond to the classification regions from Figure~\ref{fig:sims}.} \label{fig:sims2}}
\end{center}
\end{figure*}

%\vspace{1em}
\section{Predictions for Pulsar Populations}

%\newpage
%\subsection{Benefits of CS for MMA}
\label{sec:benefits}

{\refv Figure~\ref{fig:quiltdata} suggests that many pulsars in addition to PSR~B1937+21 could have high $m_{\rm cyc}$ values, most of which are not currently included in PTAs. We estimate the number of pulsars that could be successfully deconvolved in a future version of Figure~\ref{fig:quiltdata} using the example of the GBT {\refo ultrawide bandwidth} (UWB) receiver currently under construction. {\refo The GBT's UWB receiver}, similar to the UWB receiver developed by CSIRO for the Parkes Radio Telescope, spans the 0.7--4\,GHz range.

We simulated\footnote{Simulation details and results are
  available at
  \url{http://www.aoc.nrao.edu/~tcohen/research/popsynth.shtml}} the
full Galactic MSP population using the \texttt{PsrPopPy} library
\citep{blrs14} with the expected UWB system temperatures and
gains across the bandwidth, ${\mathcal T_{\rm obs}}$ = 15\,min (corresponding to typical
DISS timescales), and assuming a Gaussian pulse profile shape. {\refoo The pulsars had the requirement that $\tau_{\rm s}<P/2$}, because more highly scattered pulsars are unlikely to be
detected in {\refo untargeted} surveys. We also assumed that any pulsar with a
large value of $m_{\rm cyc}$ would have sufficiently high S/N to be
detected in future surveys.  Finally, we calculated the rms timing residual, $\sigma_{\rm TOA}$, expected \citep{lmc+18} for two cases: 1) traditional
processing, in which scattering is not mitigated for any MSP, and 2)
CS processing, in which we set $\tau_{\rm s} = 0$ for any MSP with
$m_{\rm cyc} \geq 10$ at {\refu 0.8\,GHz (equivalent to completely mitigating all IISM
noise at a typical NANOGrav timing frequency). We used {\refoo $m_{\rm cyc, full}$} from Equation~\ref{eqn:mopt}, {\reft because we assume that the high-resolution dynamic spectra from CS will resolve scintles, enabling the usage of a bandwidth over which we would measure $m_{\rm cyc}$ that varies with $\tau_{\rm s}$}. The timing rms due to variations in the IRF is assumed to be {\refo 50\% of $\tau_{\rm s}$ {\refm \citep{lmc+18}}}. {\reft The quantity $\sigma_{\rm TOA}$ also includes the rms from timing residuals from phenomena such as long-term DM variations.}}

The results are shown in Figure\ \ref{fig:sims}.  We find that $\sim$40
  MSPs that would otherwise not be of PTA-quality
    (i.e., $\sigma_{\rm TOA} > 1$\,$\us$; the pulsars shown in green) become PTA-quality
    with the use of CS and the GBT UWB receiver.  This is compared
to 45 MSPs that are always of PTA-quality using the UWB receiver, even
without the use of CS (the pulsars shown in blue).  These population simulations suggest that
CS deconvolution could double the number of PTA-quality MSPs, using the {\refv example of the GBT as one of many telescopes undergoing a receiver upgrade}.
We also find that the mean $\sigma_{\rm TOA}$ improves from $\sim$45\,$\us$ without CS to $\sim$\,2$\us$ with CS. {\refm We note that this simulated population includes pulsars similar to current NANOGrav pulsars as well as pulsars either not yet discovered or not yet included as PTA pulsars due to high scattering and/or low S/N. The off-diagonal blue points in Figure~\ref{fig:sims} likely represent pulsars not yet discovered (see Section~\ref{sec:theend}). The blue points along the diagonal represent {\refo a population similar to} NANOGrav pulsars, both discovered and yet-to-be-discovered, for which CS deconvolution does not improve timing.}

Several caveats about the cyclic merit values used to create Figure~\ref{fig:sims} bear mentioning. {\refu Because detailed profile information is difficult to extract from the population simulations, we only use the $a_1$ term in Equation~\ref{eqn:mopt}, which may \emph{underestimate} the cyclic merit of each pulsar in the simulated population, and thus the number that would improve to PTA quality. On the other hand, we also assumed that $m_{\rm cyc}>10$ implies complete scattering removal; the true degree of removal will depend on the details of the timing pipeline used (see Section~\ref{sec:theend}). The threshold $m_{\rm cyc}$ might in practice also be larger than 10; all we can say definitively is that $m_{\rm cyc} \gg 1$ for scattering removal {\refn (with motivation for a threshold of 10 from our data in Section~\ref{subsec:comparemerit})}. Although we calculated $\sigma_{\rm TOA}$ at {\refu 0.8\,GHz}, $m_{\rm cyc}$ may also need to be $>$10 at more than one frequency across {\refo the UWB receiver's} bandwidth in order to improve timing precision, possibly reducing the number of pulsars correctable to PTA-quality. As these caveats are difficult to quantify, we do not attempt to refine them further in this work. At the very least, the ``PTA-quality only with CS'' designation in Figure~\ref{fig:sims} suggests a \emph{possible} population of pulsars with good timing quality that has yet to be exploited. The artificial pulsars with a factor of $\sim$100 improvement are not known PTA pulsars that would improve by such a significant factor (note the smaller improvement factors for the ``Always PTA-quality'' pulsars). The pulsars with a factor of $\sim$100  improvement in $\sigma_{\rm TOA}$ are {\refo rather} those that in the absence of CS deconvolution would likely have never had a full timing solution prior to CS pulse-sharpening. The improvements in Figure~\ref{fig:sims} must be understood with this last consideration especially in mind.}

{\reft

Figures \ref{fig:sims2}, \ref{fig:hist}, and \ref{fig:gal} show some interesting properties of the predicted populations. Pulsars with the largest improvement (Figure~\ref{fig:sims2}) tend to be those with large $\tau_{\rm s}$ values but also narrow intrinsic pulse widths. The probability density function of pulse widths {\refs is shown in Figure \ref{fig:hist}. {\refs We measured width by $W_{\rm eff}/P$, where $1/W_{\rm eff}$ }is the mean-squared derivative of the pulse profile {\refo (\citealt{1983ApJS...53..169D}, \citealt{lcc+16})}. The distribution in Figure~\ref{fig:hist} tends} toward narrower pulse widths for the ``PTA-quality only w/CS'' sub-population, suggesting that their narrow intrinsic widths in particular will enable better timing. While these have notably narrow pulse widths, the lognormal distribution from which widths are drawn matches the {\refo width distribution from the} NANOGrav 12.5-year dataset. As Figure \ref{fig:gal} shows, the distribution of the ``PTA-quality only w/CS'' pulsars in Galactic coordinates tends toward the Galactic center, with the highest timing residual rms improvements corresponding to the most distant pulsars. Bright, intrinsically narrow, and highly scattered pulsars may very well acquire low timing rms values by means of CS.}

\begin{figure*}
\begin{center}
%\vspace{-10em}
\resizebox{1.7\columnwidth}{!}{\includegraphics{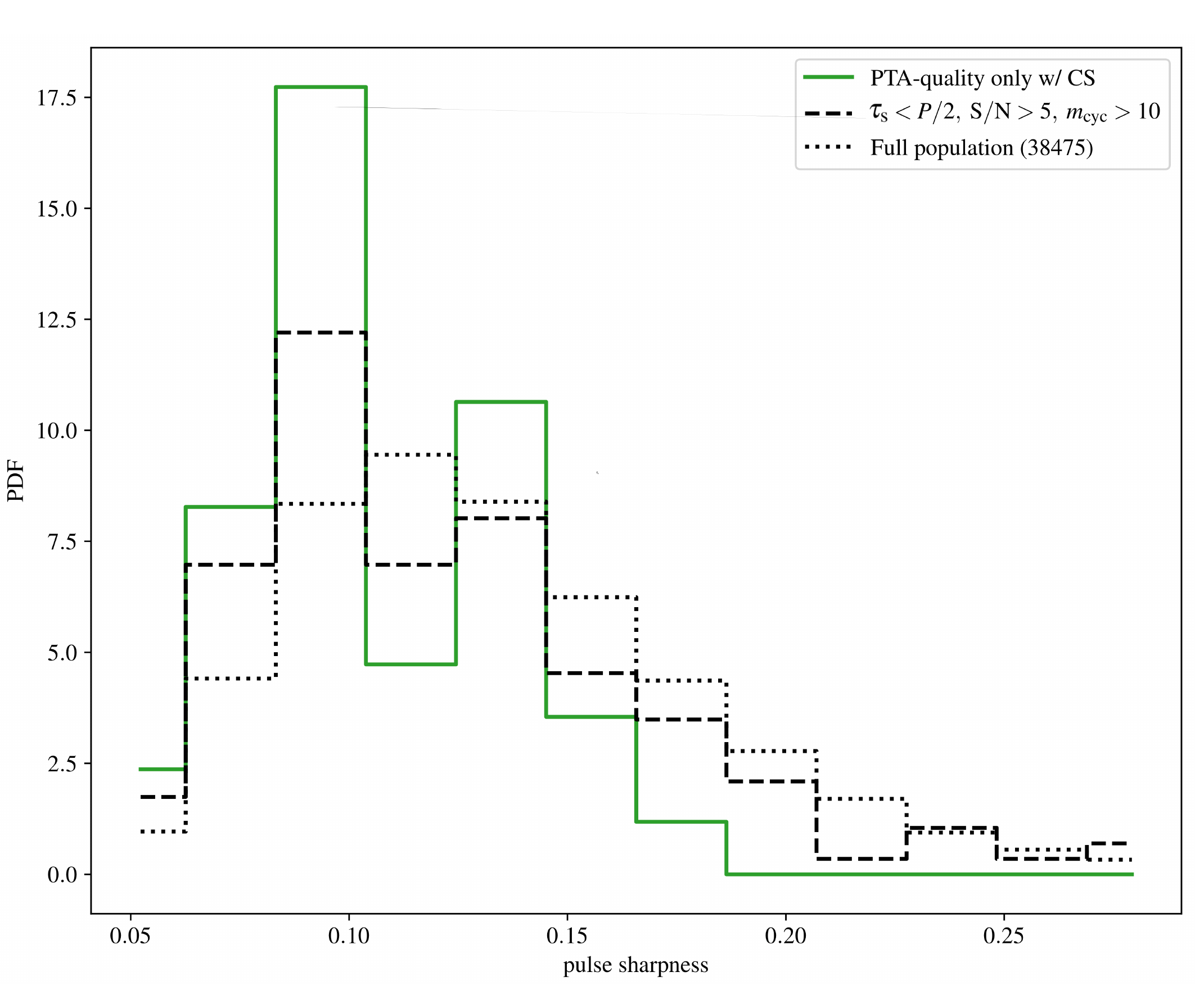}}
%\vspace{-10em}
\figcaption{{\reft Probability density functions of the pulse sharpness measurements, {\refs $W_{\rm eff}/P$}, for the intrinsic (unscattered) pulse shapes from all the simulated pulsars in Figure~\ref{fig:sims}. {\refs Both the ``PTA-quality only w/CS'' and the entire population of simulated} pulsars have sharper intrinsic pulses, enabling good timing quality if CS deconvolution is applied.} \label{fig:hist}}
\end{center}
\end{figure*}
%\vspace{-15em}
\begin{figure*}
\begin{center}
\resizebox{1.6\columnwidth}{!}{
\includegraphics[width=\columnwidth]{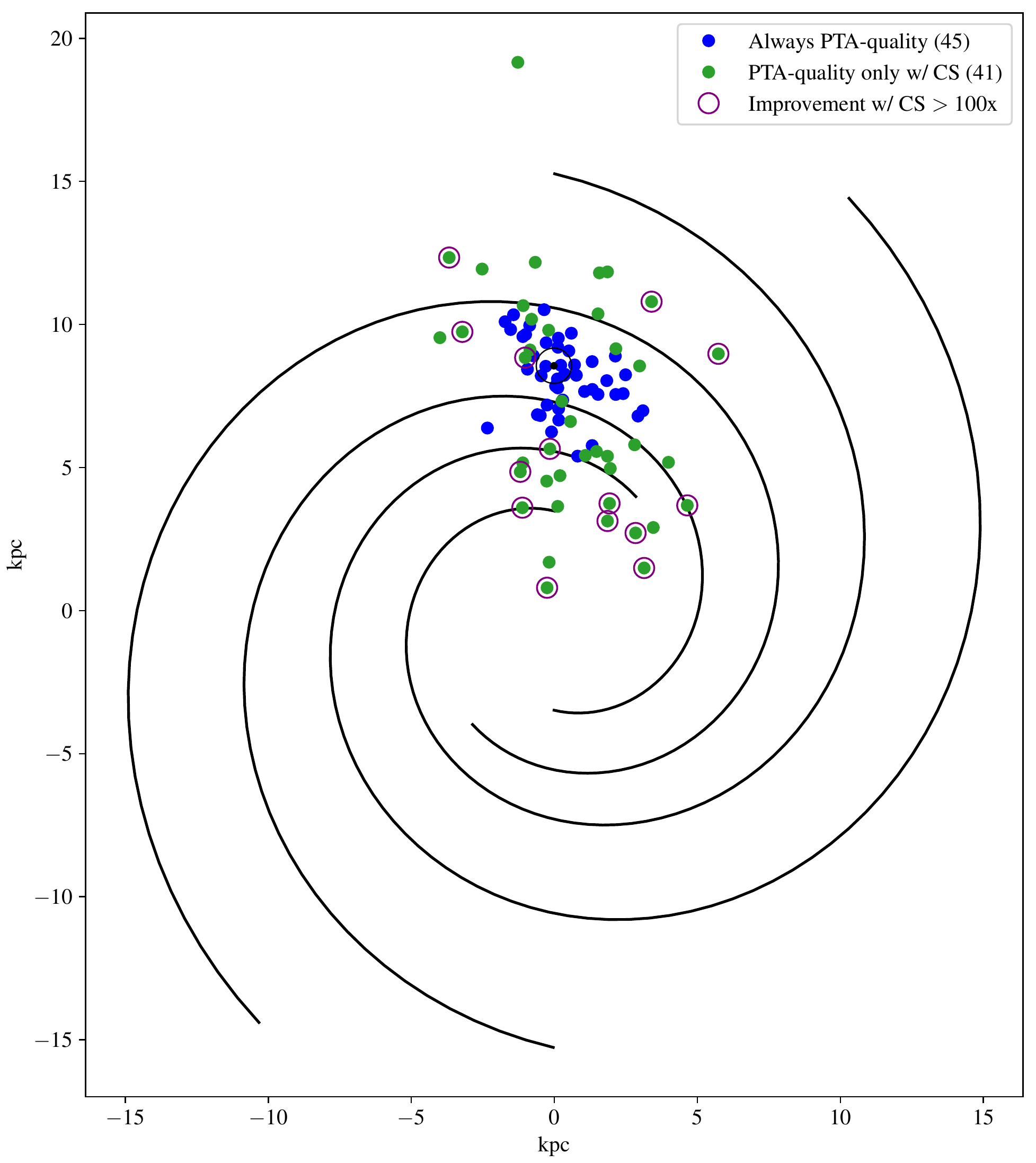}
}
\figcaption{{\reft Galactic spatial distribution of simulated pulsar set from Figure~\ref{fig:sims}. The solar system is the black circled dot. Those pulsars acquiring PTA-quality after CS deconvolution tend to be more distant than those PTA pulsars with no or small timing improvements after CS deconvolution. The small subset with extremely large timing improvements are the most distant of the ``PTA-quality only w/CS'' pulsars, consistent with their significant scattering from Figure~\ref{fig:hist}.} \label{fig:gal}}
\end{center}
\end{figure*}

}
\vspace{-0.25em}
\section{Conclusions and Next Steps}
\label{sec:theend}

{\refv To reiterate, we expect several possible CS deconvolution benefits for PTA pulsars: 
\begin{itemize}
    \item  \emph{Improved timing precision of current PTA pulsars} through scattering correction at some or all radio frequencies across a wide bandwidth. For any particular pulsar currently with PTA-quality timing precision, the timing correction may be one of many small second-order corrections, but at least for some current PTA pulsars, a S/N improvement with a new receiver and/or telescope may enable a more significant correction (blue points in Figure~\ref{fig:sims}, allowing a movement to the right in Figure \ref{fig:quiltdata}).

    \item \emph{Future pulsar discoveries}, likely to be more distant and highly scattered, that can \emph{become} PTA-quality through {\refn removal of scattering variations.} {\reft CS deconvolution also applies to known pulsars currently with low timing precision, IISM-deconvolved to have PTA-quality timing precision.}

    \item \emph{Enabling observations at lower frequencies} not presently used for pulsar timing where pulsars are intrinsically brighter, given {\reff the approximate relationship} $\tau_{\rm s} \propto \nu^{-4.4}$. {\refoo Low-frequency telescopes and instruments could become more important for pulsar timing.} %For pulsars already timed, adding the lower frequencies enables ultrawide observing.}
    
\end{itemize}
A procedure for using CS to improve pulsar timing precision is described in \citet{2015ApJ...815...89P}. The approach uses a timing model, fitting the WDS-recovered IRF with a one-sided exponential. The results are encouraging, although as the paper states, work still remains to implement a full pulse-sharpening procedure. Our present work demonstrates that the pulse-sharpening itself will likely be effective if $m_{\rm cyc} > 10$ for a particular pulsar; demonstration of a full timing pipeline involving CS deconvolution is a work in progress.

Several additional procedures will need to be incorporated into a full CS timing pipeline.} {\refxxx Scattering correction methods are most relevant to low radio frequencies.} Because high timing precision requires a DM measurement, which in turn requires high precision {\refxxx across a wide range of} frequencies, scattering variations (and thus the changing IRF) would be highly covariant with any DM variations. If scattering variations are significant enough, {\refbfr then scattering delays can be partially absorbed into the DM measurement.} Such an effect would, of course, represent a real source of error not otherwise taken into account, and {\refbfr CS} would be useful for characterizing that error, but improved timing precision is not guaranteed. If scattering is high enough, then DM measurements could become {\refbfr contaminated {\refz \citep{2016ApJ...818..166L}}}. {\refv For most pulsars, the scintillation timescale, the timescale on which IRF realizations change, is significantly less than the typical timescale for DM variations, reducing the possibility of covariance.} 

 {\refb For PSR~B1937+21, CS deconvolution is unsuccessful only for the 1.4\,GHz data presented here; at 430\,MHz, \citet{2011MNRAS.416.2821D} and \citet{2013ApJ...779...99W} successfully obtained an IRF. Being heavily subject to red noise {\refxxx (\citealt{1994ApJ...428..713K}, \citealt{2017ApJ...834...35L}, \citealt{2016ApJ...819..155L})}, PSR~B1937+21 is not a useful contributor for {\refv most expected GW signals}, given the necessity for stability on decades' timescales. It remains included in the list of routinely timed {\refbfr NANOGrav} pulsars {\refoo in order to compare arrival times between telescopes, and} because of its very high short-term timing precision, making it sensitive to burst sources on timescales of approximately {\refz months} \citep{2010ApJ...718.1400F}. Such bursts could arise from nearby individual supermassive black hole binaries in highly elliptical orbits.} PSR~B1937+21's short-term GW sensitivity could be improved by routine CS deconvolution.

{\refbfr In the case of NANOGrav, the PTA's pulsars typically have $\tau_{\rm s}$ values $<$10\,$\mu$s {\refz (\citealt{2016ApJ...818..166L}, \citealt{turner20})}, corresponding to low cyclic merit values.} {\refb This is not surprising because the pulsars were selected for high timing precision.} CS, however, remains an effective technique for obtaining extremely fine frequency resolution, and should still be performed as a routine data product. {\refbfr Finer frequency resolution helps to better excise radio frequency interference {\refo (RFI)} {\refo using the RFI itself as the cyclostationary ``signal''}, and can also provide higher frequency-resolution dynamic spectra, either for IISM studies or as a diagnostic tool for timing. For example, highly scattered epochs could be measured with fine-resolution cyclic periodograms ({\refw as demonstrated by} \citealt{2014ApJ...790L..22A}) and then appropriately down-weighted in a GW detection pipeline. Preliminary high-resolution dynamic spectra using CS were also created as part of the IPTA's 24-hour global campaign on PSR~J1713+0747 \citep{2014ApJ...794...21D}.}

{\refv The practical difficulty of CS as a routine data product is the large volume of baseband data. Real-time CS backend receivers are currently being pursued in order to create cyclic spectra and/or IISM-deconvolved timing data as a direct data product, so that baseband data will not need to be saved for post-processing.}

{\refv Within the current 12.5-year NANOGrav dataset (\citealt{alam20a}, \citealt{alam20b}), PSR~J1903+0327 is an example of a high-DM NANOGrav pulsar that would likely  benefit from such a routine procedure. {\reff Using NANOGrav 12.5-year dataset profiles, values from the ATNF pulsar catalog\footnote{\url{https://www.atnf.csiro.au/people/pulsar/psrcat/}}, and the assumption of increasing flux at low frequencies with a $-1.4$ powerlaw dependence, we estimate $m_{\rm cyc}$~$\sim$4 at 820\,MHz for PSR~J1903+0327. The true scattering timescale has not yet been estimated from resolved scintles. Improved frequency resolution with CS and/or improved telescope sensitivity could show a higher merit value. The calculation assumes an the AO sensitivity, but given the improved system temperature of the UWB receiver, this is still a pulsar worth exploring with the upgraded GBT.}  {Other distant, high-DM pulsars for which good non-NANOGrav timing solutions already exist could be added to the array.} However, {\refo these added pulsars must not be scattered so highly that low-frequency scintles are unresolvable} \citep{2010arXiv1010.3785C}; in the NANOGrav 12.5-year dataset scattering measurements \citep{turner20}, all pulsars either have $\tau_{\rm s}<1\us$ at 820\,MHz or have unresolvable scintles at {\refs the observed frequencies}. {\refu Routine, fine-frequency dynamic spectra from CS may resolve {currently unresolved scintles for PTA pulsars such as} PSR~J1903+0327, after which, CS pulse-sharpening may be applicable.}}

Other possible benefits of CS deconvolution are likely to occur for other telescopes not yet mentioned \citep{dolchswarm}. The Canadian HI Mapping Experiment (CHIME; \citealt{2014SPIE.9145E..22B}, \citealt{ng20}), although primarily searching for high-redshift baryon acoustic oscillations, {\refo observes} transiting pulsars daily from 400\,MHz to 800\,MHz, typical frequencies in which $\tau_{\rm s}$ is large enough that CS deconvolution could be successfully applied on many pulsars. Combined with the high cadence, CHIME's {\refxxx sensitivity to short-timescale GW bursts} would improve with routine CS deconvolution. The telescope is not steerable, but a pulsar-specialized backend receiver {\refz extracts} pulsar timing data from the large volume of information constantly obtained at the instrument's many receivers \citep{2020arXiv200805681C}. CHIME {\refo can obtain} high-time-sampled, low-frequency data as well as make discoveries of pulsars in its declination range. The future Square Kilometer Array, and the current pathfinders - Australian Square Kilometre Array Pathfinder (ASKAP; \citep{2009ASPC..407..446J}) and MeerKAT \citep{2013SPIE.9008E..0PG} - are expected to discover a significant fraction of all the visible pulsars in the Galaxy \citep{2004NewAR..48.1413C}.  A Next-Generation Very-Large Array (ngVLA; \citealt{2018arXiv181006594N}) has also been proposed for the northern hemisphere. Both telescopes are expected to have a factor of {\refy $\sim$10} improvement in sensitivity over {\refv AO, improving the $m_{\rm cyc}$ for many currently fainter pulsars}. As already discussed in Section~\ref{sec:benefits}, the GBT UWB receiver will enable simultaneous, highly sensitive pulsar observations across 3.3\,GHz. In Figure~\ref{fig:quiltdata}, these future telescopes or upgrades would move the example pulsars {\refy at least an order} of magnitude to the right on the diagram. {\refoo As a low-$\sigma_{\rm TOA}$ example}, PSR J1713+0747 would be located {\refy in the upper-right region} of the plot {\refoo instead of the upper left}. While CS deconvolution would require many computing nodes across a large bandwidth, such computing resources will be a part of the full SKA or ngVLA. {\refbfr With these telescopes, many pulsars will have S/N similar to that of PSR~B1937+21 now.} These changes could heavily populate the diagnostic diagram in this paper with pulsars that are high S/N enough and/or highly scattered enough for feasible routine CS deconvolution. 

{\refy {\refz In addition to the successful application of CS to LOFAR data \citep{2014ApJ...790L..22A}}, other low-frequency radio telescopes (e.g., 200\,MHz and below) could benefit from routine CS data products, for the sake of both fine channelization data products and for deconvolution. While these frequencies do not usually allow for precision timing, the significant scattering tails present could be utilized vis-a-vis CS deconvolution to significant improve timing precision, {\refn or to arrive at stable timing solutions otherwise unavailable before CS deconvolution}. {\refoo The Long-Wavelength Array (LWA), recently expanded to two stations with a significant sensitivity increase, is a relevant example.} The LWA is already monitoring many pulsars {\refw \citep{2015ApJ...808..156S}, {\refxxx several} of which are currently NANOGrav MSPs\footnote{\url{https://lda10g.alliance.unm.edu/}}}. Future low-frequency telescopes such as SKA-low in the southern hemisphere and the proposed LWA-Swarm telescope (\citealt{2018JAI.....750006D}, \citealt{swarm}) {\refo may} have sensitivities for a significant S/N boost in Figure~\ref{fig:quiltdata}. The Expanded LWA (ELWA; \citealt{2017arXiv170800090T}, \citealt{ruan20}) already uses recently installed 50--80\,MHz dipoles on the VLA together with the two LWA stations. {{\refoo Even if low-frequency timing does not become feasible}, these low-frequency telescopes aided with CS can complement timing observations with DM measurements and with monitoring of {\refoo CS-resolved} secondary spectra {\refoo (the parabolic 2D FTs of dynamic spectra containing scattering structure information)} as done with PSR~B1937+21 in \citet{2013ApJ...779...99W}.} Finally, CS has been successfully applied to data from the VLA Low Band Ionospheric and Transient Experiment (VLITE; \citealt{2016ApJ...832...60P}) project in order to provide very high-frequency resolution for RFI excision (M.~Kerr, private communication). For the transient search aspect of VLITE, the improved frequency resolution {\refyy of periodic RFI \citep{2011MNRAS.416.2821D}, and therefore the improved quality of RFI excision,} could aid the search for fast radio bursts or other radio transients.} {\refs The Breakthrough Listen project, along these lines, has searched for technosignature periodicities in association with FRBs \citep{10.1093/mnras/stz958}, by creating cyclic spectra, shifting at multiples of small time intervals and seeing if power appears in harmonics of a particular period.} Beyond transient radio astronomy, as in radio spectral line observations, routine CS would also help excise RFI.

{\refu CS may also serve as an important pulsar searching tool. } {\refxxx The presence of scattering in a CS phase slope could serve as an additional pulsar searching metric, along with trial DMs and periods, for an astrophysically pulsating signal. Searches near the highly scattered Galactic center might particularly benefit {\refo from CS as a routine and/or real-time data product}.}
%\newline
%\footnote{\url{https://lda10g.alliance.unm.edu/}}
%\vspace{-4em}
\section{Acknowledgments}
%\vspace{-1em}
{\reff We thank the referee for the many helpful suggestions that have been incorporated into this paper.} {\refbfr {\refzz We thank J. M. Cordes, {\refu T. T. Pennucci,} {\refyy M. Kerr,} and the NANOGrav collaboration for helpful discussions and suggestions,} {the Noise Budget Working Group in particular}. We thank the AO staff for their assistance during the observations reported in this work, in particular P. Perillat, J. S. Deneva, H. Hernandez, and the telescope operators.  {\refw We are grateful for the use of the Bowser computing cluster at WVU with the assistance of N.~E.~Garver-Daniels.} {\reft From Hillsdale College, TD gratefully acknowledges computing resources from the Department of Physics, summer leave funds, and professional development funds for conducting this research.} The authors were partially supported through the National Science Foundation (NSF) PIRE program award number 0968296. Members of the NANOGrav Collaboration {\refy (TD, DS, RL, PD, ML, MM)} acknowledge support from the NSF Physics Frontiers Center (PFC) award number 1430284. {\reft TD acknowledges NANOGrav seed-funding for meetings where this research was conducted.} AO is a facility of the NSF operated under cooperative agreement by UCF in alliance with Yang Enterprises, Inc. and Universidad Metropolitana. {\refv The Green Bank Observatory (GBO) is a facility of the NSF operated under cooperative agreement by Associated Universities, Inc.} {\refv The GBT UWB receiver project is funded in part by the Gordon and Betty Moore Foundation through Grant GBMF7576 to Associated Universities Inc. to support the work of the GBO and the NANOGrav PFC.}} TD and MTL are supported by an NSF Astronomy and Astrophysics Grant (AAG) award number 2009468. This work made use of NASA's ADS and the arXiv.org preprint service.%This work made use of NASA’s ADS Abstract Service and the arXiv.org preprint service.}}%This work used ADS and arXiv.org.}} 
%
%\newpage
%{\refxxx
%\vspace{3em}
\appendix
\label{sec:appendi-cite-us}
%\vspace{-2em}
\section{Derivation of the Cyclic Merit}
According to Equation \ref{eqn:phaseslope}, the phase of the cyclic spectrum has two contributions: one term induced by the derivative of the phase of the IISM transfer function and the other intrinsic to the pulse profile. By the shift theorem in Fourier analysis, a translation in the time domain induces a phase slope in the frequency domain. Since the time-domain impulse response function is characterised by a delay $\tau_{\rm s}$, we can estimate the phase derivative of the transfer function by $-2\pi \tau_{\textrm{s}}$,
\begin{align}
\Phi_{S_E}(\nu, \alpha_k)
&= \frac{k}{P}\frac{\textrm{d}\Phi_H}{\textrm{d}\nu}+ \phi_{S_x}(\alpha_k)\\
&\approx -2\pi\frac{ k\tau_{\textrm{s}}}{P} + \phi_{S_x}(\alpha_k)\\
&\approx k\, b + \phi_{S_x}(\alpha_k),
\label{eqn:phaseslope2b}
\end{align}
where $b=2\pi \tau_{\rm{s}}/P$ is the average cyclic phase slope, also equal to the expected phase at $k=1$.

The spectrum of the intrinsic pulse profile can be determined accurately by the WDS algorithm, since the Hessian is diagonal with respect to the set of parameters describing the profile \citep{2013ApJ...779...99W}. Hence, in general, the second term in Equation~\ref{eqn:phaseslope2b} does not complicate the phase retrieval process. At larger modulation frequencies, the CS amplitude is lost to the noise as in Figure~\ref{fig:cycamp} {\refv near $\alpha_k$ = 40}, which causes a transition to incoherence in phase as in Figure \ref{fig:cycphase}. We can estimate the phase noise in a cyclic spectrum at the $k^{th}$ harmonic to be {\refv
\begin{align}
\Delta\Phi_k \equiv \sigma_{\textrm{noise}}/{A_k}
\end{align}}\noindent
where $A_k$ is the frequency-averaged amplitude of the cyclic spectrum at harmonic $\alpha_k$. By a weighted linear fit, the phase slope $b$ can be estimated by
\begin{align}
    b = \frac{1}{\Delta_w}\sum_k{w_k\, k\, \Phi_E(k)},
\end{align}
where the least-square weights are $w_k=1/[\delta\Phi_E(k)]^2$, and $\Delta_w \equiv \sum_k w_k\, k^2$ does not depend on the measured $\Phi_E(k)$ values.
Then, the uncertainty in the phase slope $b$ is
\begin{align}
    \delta b &= \frac{1}{\Delta_w}\sqrt{\sum_k{w_k^2\, k^2\, [\delta\Phi_E(k)]^2}}\\ 
    &=\frac{1}{\Delta_w}\sqrt{\sum_k \frac{k^2} {[\delta\Phi_E(k)]^2}},
\end{align}
where all the sums are from 1 to $N_{\rm max}/2 + 1$ and where $N_{\rm max}$ is the number of time samples in the folded profile.
This can be further simplified by expressing the uncertainty of the cyclic spectrum phase in terms of observational parameters {\refv from the radiometer equation:
\begin{align}
    \delta\Phi_E(k) \approx \Delta\Phi_k = \frac{\Delta\Phi_0}{a_k}  &= \frac{1}{a_k}\frac{T_{\rm sys}}{g S_0\sqrt{B{\mathcal T_{\rm obs}}}}.
\end{align}}\noindent 
Here, $T_{\rm sys}$ is the system temperature, $g$ is the telescope gain (units: K/Jy), $S_0$ is the pulsar continuum flux (single polarization), $B$ is the bandwidth over which the phase information is integrated {\refv (ideally channelized to the diffractive bandwidth, $\Delta\nu_d$), and $\mathcal T_{\rm obs}$ is the  observing time in the integration.
Taken together, this allows us to define a  quality metric, the \emph{cyclic merit},} for cyclic spectroscopy phase retrieval, which we express as the signal-to-noise ratio of the cyclic spectrum phase
\begin{align}
    m_{\rm cyc} = \frac{b}{\delta b} = \frac{2\pi\tau_{\rm s}}{P} \frac{g S_0}{T_{\rm sys}} \sqrt{B{\mathcal T_{\rm obs}} \sum_k k^2 a_k}.
    \label{eq:cm}
\end{align}
{\refv The above metric can then be written in terms of the folded pulse profile signal-to-noise ratio $S/N$}

\begin{equation}
    m_{\rm cyc} =  \frac{2\pi\tau_{\rm s} W_e}{P^2} (S/N) \sqrt{ \sum_k k^2 a_k},%m_{\rm cyc} =  \frac{2\pi\tau_{\rm s} W_e}{P^2} (\mathrm{S/N}) \sqrt{ \sum_k k^2 a_k
    \label{eq:cm2}
\end{equation}
where $W_e$ is the equivalent width of the pulse profile, and
\begin{equation}
(S/N) \equiv \frac{gS_0}{T_{\rm sys}}\frac{P}{W_e}\sqrt{B{\mathcal T_{\rm obs}\left(\frac{W_e}{P}\right)}} = \frac{gS_0}{T_{\rm sys}}\sqrt{B{\mathcal T_{\rm obs}\left(\frac{P}{W_e}\right)}}.
%(\mathrm{S/N}) \equiv \frac{gS_x}{T_{\rm sys}}\frac{P}{W_e}\sqrt{B{\mathcal T_{\rm obs}}}.
\label{eqn:mcycsn}
\end{equation}
Here we have assumed that the profile has been smoothed to a time resolution of $W_e$ for an optimal trade-off between pulse sharpness and $S/N$.
%\footnote
{{\reff In principle, the smoothing should be done to a time resolution of the sharpness width, $W_s \equiv \int{dt |U'(t)|^2}$, where $U(t)$ is a unit amplitude average pulse profile, and the prime indicates a time derivative  \citep{2010arXiv1010.3785C}. In practice, the quantitative difference between $W_e$ and $W_s$ is expected to be small for realistic pulse profiles.}}

%\noindent 
Equation~\ref{eq:cm} is the cyclic merit for determining the {\em average} phase slope, which is a useful estimator of $\tau_s$.
If a full reconstruction of the transfer function is required, we must estimate the cyclic spectrum phase over a diffractive bandwidth.
This requires setting $B=\Delta\nu_d \approx 1/(2\pi\tau_{\rm s})$ in Equation \ref{eq:cm} from which we obtain a cyclic merit for a full reconstruction of the transfer function

\begin{equation}
    m_{\rm cyc,full} =  \frac{g S_0}{P T_{\rm sys}} \sqrt{{2\pi \tau_s \mathcal T_{\rm obs}} \sum_k k^2 a_k},
\label{eqn:mopt}
\end{equation}
where we see a square-root dependence on the scattering delay.
\newpage
\bibliographystyle{apj}
%\bibliography{apjmnemonic,manuscript_cs_simulations}
\bibliography{manuscript_cs_simulations}%,scintillation}{}
\bibliographystyle{aasjournal}
%\printbibliography
\end{document}

%% file: authors.tex
\author[0000-0001-8885-6388]{Timothy Dolch}
\affiliation{Department of Physics, Hillsdale College, 33 E. College Street, Hillsdale, MI 49242, USA}
\affiliation{Eureka Scientific, Inc.  2452 Delmer Street, Suite 100, Oakland, CA 94602-3017}

\author[0000-0002-1797-3277]{Daniel R. Stinebring}
\affiliation{Department of Physics and Astronomy, Oberlin College, Oberlin, OH 44074, USA}

\author{Glenn Jones}
\affiliation{Rigetti Computing, Inc., 775 Heinz Ave. Berkeley, CA 94710, USA}
\affiliation{Columbia Astrophysics Laboratory, Columbia University, NY 10027, USA}

\author[0000-0001-9027-4184]{Hengrui Zhu}
\affiliation{Department of Physics and Astronomy, Oberlin College, Oberlin, OH 44074, USA}

\author[0000-0001-5229-7430]{Ryan S. Lynch}
\affiliation{Green Bank Observatory, P.O. Box 2, Green Bank, WV 24944, USA}

\author[0000-0001-7587-5483]{Tyler Cohen}
\affiliation{Department of Physics, New Mexico Institute of Mining and Technology, 801 Leroy Place, Socorro, NM 87801, USA}

\author[0000-0002-6664-965X]{Paul B. Demorest}
\affiliation{National Radio Astronomy Observatory, 1003 Lopezville Rd., Socorro, NM 87801, USA}

\author[0000-0003-0721-651X]{Michael T. Lam}
\affiliation{School of Physics and Astronomy, Rochester Institute of Technology, Rochester, NY 14623, USA}
\affiliation{Laboratory for Multiwavelength Astronomy, Rochester Institute of Technology, Rochester, NY 14623, USA}

\author[0000-0002-2034-2986]{Lina Levin}
\affiliation{Jodrell Bank Centre for Astrophysics, University of Manchester, Manchester, M13 9PL, United Kingdom}

\author[0000-0001-7697-7422]{Maura A. McLaughlin}
\affiliation{Department of Physics and Astronomy, West Virginia University, P.O. Box 6315, Morgantown, WV 26506, USA}
\affiliation{Center for Gravitational Waves and Cosmology, West Virginia University, Chestnut Ridge Research Building, Morgantown, WV 26505, USA}

\author[0000-0002-4828-0262]{Nipuni T. Palliyaguru}
\affiliation{Department of Physics \& Astronomy
Texas Tech University, Lubbock, TX 79409, USA}

%% file: abstract.tex
\newcommand{\refbfr}{}

\newcommand{\refb}{}
\newcommand{\refx}{}
\newcommand{\refxx}{}
\newcommand{\refq}{}
\newcommand{\refw}{}
\newcommand{\refxxx}{}
\newcommand{\refy}{}
\newcommand{\refz}{}
\newcommand{\refzz}{}
\newcommand{\refyy}{}
\newcommand{\refv}{}
\newcommand{\refu}{}
\newcommand{\reft}{}
\newcommand{\refs}{}
\newcommand{\refm}{}
\newcommand{\refn}{}
\newcommand{\refo}{}
\newcommand{\refoo}{}
\newcommand{\reff}{}

\begin{abstract}

%The deconvolution methodology outlined in recent cyclic
%spectroscopy (CS) papers (Demorest 2011; Walker, Demorest, and van
%Straten 2013) can be used to measure the interstellar
%scattering (ISS) of millisecond pulsars. Correcting times-of-arrival
%for ISS is of interest for detecting gravitational waves with pulsar
%timing arrays. We present an analysis of simulated data to test the
%performance of CS deconvolution over a range of scattering and
%signal-to-noise parameters. We show that the inherent biases and
%scatter of the pulse-broadening function recovery must be taken into
%account when interpreting pulsar data through CS. Consequently, we set
%boundary conditions in the space of pulsar and ISS parameters within
%which the CS deconvolution technique can be effectively applied.

Radio pulsar signals are significantly perturbed by their propagation through the ionized interstellar medium. In addition to the frequency-dependent pulse times of arrival due to dispersion, pulse shapes are also distorted {\refxxx and shifted}, having been scattered by the inhomogeneous interstellar plasma, affecting pulse arrival times. Understanding the degree to which scattering affects pulsar timing is important for gravitational wave detection with pulsar timing arrays (PTAs), which depend on the reliability of pulsars as stable clocks with an uncertainty of $\sim$100\,ns or less over $\sim$10\,yr or more. Scattering can be {\refb described} as a convolution of the intrinsic pulse shape with an impulse response function {\refb representing the effects of multipath propagation.} {\refoo In previous studies, the technique of cyclic spectroscopy has been applied to pulsar signals to deconvolve the effects of scattering from the original emitted signals,} increasing the overall timing precision. We present an analysis of simulated data to test the quality of deconvolution using cyclic spectroscopy over a range of parameters characterizing interstellar scattering and pulsar signal-to-noise {\refb ratio}. {\refw We show that cyclic spectroscopy is most effective for high-S/N and/or highly scattered pulsars.} We conclude that cyclic spectroscopy could play an important role in scattering correction to distant populations of highly scattered pulsars not currently included in PTAs. {\refoo For future telescopes and for current instruments such as the {\refv Green Bank Telescope upgraded with the {\refo ultrawide bandwidth (UWB) receiver, cyclic spectroscopy could potentially double the number of PTA-quality pulsars.}}}%, as well as {\refy {\refo low-frequency ($<$200\,MHz) radio telescopes}.}%as well as in the era of instruments such as the }Square Kilometer Array (SKA), the Canadian HI Mapping Experiment (CHIME), {\refv the Green Bank Telescope upgraded with the Ultra-Wide Band (UWB) receiver}, {\refxxx the Arecibo Observatory upgraded with the Advanced L-band Phased Array Camera for Arecibo (ALPACA)} and the Next-Generation Very Large Array (ngVLA), {\refy as well as {\refw in pulsar astrophysics with low-frequency ($<$200\,MHz) radio telescopes generally}.}

\end{abstract}